\documentclass[twocolumn,showpacs,amsmath,amssymb,superscriptaddress]{revtex4-1}
\usepackage{dcolumn}
\usepackage{color}
\usepackage{graphicx}
\usepackage{amsmath}
\usepackage{multirow}
\usepackage{bm}
\usepackage{url}

\begin{document}

\title{Shape-phase transitions in odd-mass $\gamma$-soft nuclei with
mass $A\approx 130$}

\author{K.~Nomura}
\affiliation{Physics Department, Faculty of Science, University of Zagreb, 10000 Zagreb, Croatia}
\affiliation{Center for Computational
Sciences, University of Tsukuba, Tsukuba 305-8577, Japan}

\author{T.~Nik\v si\'c}
\author{D.~Vretenar}
\affiliation{Physics Department, Faculty of Science, University of Zagreb, 10000 Zagreb, Croatia}

\date{\today}

\begin{abstract}

Quantum phase transitions between competing equilibrium shapes of nuclei
 with an odd number of nucleons are explored using a microscopic framework
 of nuclear energy density functionals and a fermion-boson coupling model. 
The boson Hamiltonian for the even-even core nucleus, as well as the
 spherical single-particle energies and occupation 
 probabilities of unpaired nucleons, are completely determined by a  
 constrained self-consistent mean-field calculation for a specific
 choice of the energy density functional and pairing interaction. 
Only the strength parameters of the particle-core coupling have to be
 adjusted to reproduce a few empirical low-energy spectroscopic properties of
 the corresponding odd-mass system. 
The model is applied to the odd-A
 Ba, Xe, La and Cs isotopes with mass $A\approx 130$, for which the
 corresponding even-even Ba and Xe nuclei present a typical case 
of $\gamma$-soft nuclear potential. 
The theoretical results reproduce the experimental low-energy excitation spectra and electromagnetic
 properties, and confirm that a phase transition between nearly spherical and
 $\gamma$-soft nuclear shapes occurs also in the odd-A systems. 

\end{abstract}


\maketitle


\section{Introduction}


In many areas of physics and chemistry, quantum phase transitions
(QPT) present a prominent feature of strongly-correlated many-body systems
\cite{carr-book}. 
In atomic nuclei, in particular, a QPT occurs between competing ground-state shapes
(spherical, axially-deformed, and $\gamma$-soft shapes) as a function of a
non-thermal control parameter -- the nucleon number \cite{cejnar2010}. 
Even though in most cases nuclear shapes evolve gradually with nucleon
number, in specific instances, with the 
addition or subtraction of only few nucleons a shape transition occurs 
characterized by a significant change of several observables and can be 
classified as a first-order or second-order QPT. Of course, in systems with a 
finite number of particles a QPT is smoothed out to a certain extent and, 
in the nuclear case, the physical control parameter takes on only integer values. 
Therefore, the essential issue in nuclear QPT concerns the 
identification of a particular nucleus at a critical point of phase transition, and
the evaluation of observables that can be related to quantum order
parameters.

Several empirical realizations of nuclear QPT and related
critical-point phenomena have been observed in different regions of the chart of nuclides. 
In the rare-earth region, for instance, a rapid structural change 
occurs from spherical vibrational to axially-deformed rotational
nuclei, and is associated with a first-order QPT \cite{iachello2001,casten2001}. 
Evidence of a second-order QPT that occurs
between spherical vibrational and $\gamma$-soft systems has also 
been found in several mass regions, and one of the best studied cases 
are the Ba and Xe nuclei with mass $A\approx 130$. 
In particular, the isotope $^{134}$Ba has been identified
\cite{casten2000} as the first empirical realization of the E(5)
critical-point symmetry \cite{iachello2000} of a second-order QPT.

Numerous theoretical studies have explored, predicted and described 
nuclear QPTs, based on the nuclear
shell model \cite{shimizu2001,togashi2016}, nuclear density functional theory
\cite{niksic2007,li2010}, geometrical models \cite{cejnar2010}, and
algebraic approaches \cite{cejnar2009,cejnar2010}. Most of these
analyses, however, have only considered even-even nuclei. 
In these systems nucleons are coupled pairwise,  
and the low-energy excitation spectra are characterized by collective 
degrees of freedom \cite{BM,bohr1953}. 
A theoretical investigation of QPT in systems with odd $Z$ and/or $N$
can be much more complicated because one needs to consider both the collective
(even-even core) as well as single-particle (unpaired nucleon(s)) degrees of
freedom that determine low-energy excitations \cite{bohr1953}. 
Important questions that must be addressed when considering QPTs in odd-A systems
include the effect of the odd particle on the location and nature of a phase
transition, and the identification and evaluation of quantum order
parameters. QPTs in odd-mass systems currently present a very active research 
topic \cite{iachello2011a,iachello2011,petrellis2011}. 
In the last few years a number of phenomenological methods have been
developed to study QPTs in odd-A nuclei 
\cite{iachello2011,iachello2011a,petrellis2011,zhang2013a,zhang2013b,bucurescu2017}. 
Microscopic approaches, however, have not been extensively applied to QPTs in these 
systems.

Recently we have developed a new theoretical method 
\cite{nomura2016odd} for odd-mass nuclei, that is based on nuclear 
density functional theory and the particle-vibration coupling scheme. 
In this approach the even-even core is described in the framework of the
interacting boson model (IBM) \cite{IBM} using $s$ and
$d$ bosons, that correspond to collective 
pairs of valence nucleons with $J^{\pi}=0^+$ and $2^+$ \cite{OAI}, respectively, and for the
particle-core coupling the interacting boson-fermion
model (IBFM) \cite{IBFM} is used. The deformation energy surface of an 
even-even nucleus as a function of the quadrupole shape variables
$(\beta, \gamma)$, as well as the single-particle energies and occupation probabilities
of the odd nucleons, are obtained in a self-consistent
mean-field calculation for a specific choice of the nuclear energy density
functional (EDF) and a pairing interaction, and they determine the  
microscopic input for the parameters of the IBFM Hamiltonian. 
Only the strength parameters of the boson-fermion coupling terms in the
IBFM Hamiltonian have to be adjusted to low-energy data in the considered odd-A nucleus. 
In Ref.~\cite{nomura2016qpt} this method has been applied to an analysis of 
the signatures of shape phase transitions in the axially-deformed
odd-mass Eu and Sm isotopes, and several mean-field and spectroscopic
properties have been identified as possible quantum order
parameters of the phase transition.

The aim of this work is to extend the analysis of 
Ref.~\cite{nomura2016qpt} to $\gamma$-soft odd-A systems. 
In the present study we consider odd-A Xe ($Z=54$), Cs ($Z=55$),
Ba ($Z=56$) and La ($Z=57$) isotopes with mass $A\approx 130$. 
As mentioned above, the even-even nuclei of Ba and Xe in this mass region 
present an excellent example of a second-order QPT that occurs between nearly
spherical and $\gamma$-soft equilibrium shapes \cite{cejnar2010}. 
The low-lying states of the corresponding odd-A nuclei are described in terms of
the even-even cores Ba and Xe, coupled to an unpaired neutron (odd-A Ba and
Xe) or proton (odd-A La and Cs). 
Similarly to our previous work on QPTs in odd-N (Sm) and odd-Z
(Eu) nuclei \cite{nomura2016qpt}, here we consider the two possible cases that 
arise in odd-A systems: (i) the unpaired nucleon (neutron) is of the same type as 
the control parameter (neutron number) of the corresponding 
even-even boson core nuclei (the case of odd-A Ba and Xe), and (ii)
the unpaired nucleon (proton) is of different type from the control parameter (the case of odd-A La and Cs).
In general, the boson-fermion interaction will not be the same in the two cases and, therefore, 
one expects a distinct effect on the shape-phase transition that characterizes the even-even boson core.

Section~\ref{sec:model} contains a short outline of the theoretical method 
used in the present study. In Sec.~\ref{sec:results} we analyze the
deformation energy surfaces for the even-even Ba and Xe isotopes, and 
compare the calculated low-energy excitation spectra and electromagnetic properties 
of the odd-mass Ba, Xe, La, and Cs nuclei to available spectroscopic data. 
We also compute and examine quadrupole shape invariants as
signatures of shape phase transitions in odd-A $\gamma$-soft systems. 
A summary of the main results and a brief outlook for future studies are included in 
Sec~\ref{sec:summary}.


\section{Model particle-core Hamiltonian\label{sec:model}}


The IBFM Hamiltonian, used here to describe the structure of excitation spectra of odd-A nuclei, 
consists of three terms: the even-even boson-core IBM Hamiltonian $\hat H_B$, 
the single-particle Hamiltonian for the unpaired fermions $\hat H_F$, and the boson-fermion 
coupling Hamiltonian $\hat H_{BF}$. 
\begin{eqnarray}
\label{eq:ham}
 \hat H=\hat H_B + \hat H_F + \hat H_{BF}. 
\end{eqnarray}
The number of bosons $N_B$ and fermions $N_F$ are assumed to be
conserved separately and, since in the present study we only consider low-energy excitation 
spectra, $N_F=1$.
The building blocks of the IBM framework are $s$ and $d$ bosons that 
represent collective pairs of valence nucleons coupled
to angular momentum $J^\pi=0^+$ and $2^+$, respectively \cite{OAI}. 
$N_B$ equals the number of valence fermion pairs, and no distinction is  
made between proton and neutron bosons. 
We employ the following form for the IBM Hamiltonian $\hat H_B$: 
\begin{eqnarray}
\label{eq:ibm}
 \hat H_B = \epsilon_d\hat n_d + \kappa\hat Q_B\cdot\hat Q_B, 
\end{eqnarray}
with the $d$-boson number operator $\hat n_d=d^{\dagger}\cdot\tilde d$,
and the quadrupole operator $\hat Q_B=s^{\dagger}\tilde d+d^{\dagger}\tilde s +
\chi[d^{\dagger}\times\tilde d]^{(2)}$. 
$\epsilon_d$, $\kappa$, and $\chi$ are strength parameters. 
The single-fermion Hamiltonian reads $\hat
H_F=\sum_{j}\epsilon_j[a^{\dagger}_j\times\tilde a_j]^{(0)}$, 
where $a^{\dagger}_j$ and $a_j$ are the fermion creation and annihilation
operators, respectively, and
$\epsilon_j$ denotes the single-particle energy of the orbital $j$. 
For the boson-fermion coupling Hamiltonian $\hat H_{BF}$ we use \cite{IBFM}: 
\begin{eqnarray}
\label{eq:bf}
 \hat H_{BF}
=&&\sum_{jj^{\prime}}\Gamma_{jj^{\prime}}\hat
  Q_B\cdot[a^{\dagger}_j\times\tilde a_{j^{\prime}}]^{(2)} \nonumber \\
&&+\sum_{jj^{\prime}j^{\prime\prime}}\Lambda_{jj^{\prime}}^{j^{\prime\prime}}
:[[d^{\dagger}\times\tilde a_{j}]^{(j^{\prime\prime})}
\times
[a^{\dagger}_{j^{\prime}}\times\tilde d]^{(j^{\prime\prime})}]^{(0)}:
\nonumber \\
&&+\sum_j A_j[a^{\dagger}\times\tilde a_{j}]^{(0)}\hat n_d, 
\end{eqnarray}
where the first and second terms are referred to as the
quadrupole dynamical and exchange interactions, respectively. 
The third term represents a monopole boson-fermion interaction.
The strength parameters $\Gamma_{jj^{\prime}}$,
$\Lambda_{jj^{\prime}}^{j^{\prime\prime}}$ and $A_j$ can be expressed,
by use of the generalized seniority scheme, in the following
$j$-dependent forms \cite{scholten1985}: 
\begin{eqnarray}
\label{eq:dynamical}
&&\Gamma_{jj^{\prime}}=\Gamma_0\gamma_{jj^{\prime}} \\
\label{eq:exchange}
&&\Lambda_{jj^{\prime}}^{j^{\prime\prime}}=-2\Lambda_0\sqrt{\frac{5}{2j^{\prime\prime}+1}}\beta
_{jj^{\prime\prime}}\beta_{j^{\prime}j^{\prime\prime}} \\
\label{eq:monopole}
&&A_j=-A_0\sqrt{2j+1}
\end{eqnarray}
where
$\gamma_{jj^{\prime}}=(u_ju_{j^{\prime}}-v_jv_{j^{\prime}})Q_{jj^{\prime}}$
and 
$\beta_{jj^{\prime}}=(u_jv_{j^{\prime}}+v_ju_{j^{\prime}})Q_{jj^{\prime}}$,
with the matrix element of the quadrupole operator in the single-particle
basis $Q_{jj^{\prime}}=\langle j||Y^{(2)}||j^{\prime}\rangle$. 
The factors $u_j$ and $v_j$ denote the occupation amplitudes of the
orbit $j$, and satisfy the relation $u_j^2+v_j^2=1$. 
$\Gamma_0$, $\Lambda_0$ and $A_0$ are strength parameters 
that have to be adjusted to low-energy structure data. A more detailed description of the model, 
and a discussion of various approximations, 
can be found in Ref.~\cite{nomura2016odd}.

The first step in the construction of the IBFM Hamiltonian 
Eq.~(\ref{eq:ham}) are the parameters of the boson-core IBM term 
$\hat H_B$ that are determined using the mapping procedure
developed in Refs.~\cite{nomura2008,nomura2010,nomura2011rot}: the
($\beta, \gamma$)-deformation energy surface, obtained in a constrained
self-consistent mean-field calculation that also includes pairing correlations, is mapped onto
the expectation value of $\hat H_B$ in the boson condensate state
\cite{ginocchio1980}. This procedure fixes the values of the parameters
$\epsilon_d$, $\kappa$ and $\chi$ of the boson Hamiltonian $\hat H_B$. 
As in our two previous studies of Refs.~\cite{nomura2016odd} and \cite{nomura2016qpt},
the deformation energy surfaces of even-even Ba and Xe isotopes are calculated using the 
relativistic Hartree-Bogoliubov model based on the energy density 
functional DD-PC1 \cite{DDPC1}, and a separable pairing force of finite
range \cite{tian2009}. The corresponding parameters of the IBM Hamiltonian 
for the isotopes $^{128-136}$Ba and $^{126-134}$Xe are listed in
Table~\ref{tab:paraB}.

\begin{table}[hb!]
\caption{\label{tab:paraB} Parameters of the boson Hamiltonian $\hat H_B$
 ($\epsilon_d$, $\kappa$, and $\chi$) for $^{128-136}$Ba and $^{126-134}$Xe. 
 The values of $\epsilon_d$ and $\kappa$ are in units of MeV, while 
 $\chi$ is dimensionless. }
\begin{center}
\begin{tabular*}{\columnwidth}{p{2.0cm}p{2.0cm}p{2.0cm}p{2.0cm}}
\hline\hline
\textrm{} &
\textrm{$\epsilon_{d}$}&
\textrm{$\kappa$}&
\textrm{$\chi$} \\
\hline
$^{128}$Ba & 0.03 & -0.102 & -0.18 \\
$^{130}$Ba & 0.06 & -0.116 & -0.18 \\
$^{132}$Ba & 0.13 & -0.122 & -0.18 \\
$^{134}$Ba & 0.38 & -0.124 & -0.24 \\
$^{136}$Ba & 1.15 & -0.122 & -0.85 \\
$^{126}$Xe & 0.13 & -0.115 & -0.16 \\
$^{128}$Xe & 0.06 & -0.132 & -0.18 \\
$^{130}$Xe & 0.07 & -0.142 & -0.18 \\
$^{132}$Xe & 0.3 & -0.144 & -0.52 \\
$^{134}$Xe & 0.65 & -0.144 & -0.88 \\
\hline\hline
\end{tabular*}
\end{center}
\end{table}

For the fermion valence space we include all the spherical
single-particle orbitals in the proton (neutron) major shell
$Z(N)=50-82$ for the odd-A La and Cs (Ba and Xe) isotopes: $3s_{1/2}$,
$2d_{3/2}$, $2d_{5/2}$ and $1g_{7/2}$ for positive-parity states, and
$1h_{11/2}$ for negative-parity states.  
Consistent with the definition of the IBFM Hamiltonian, the spherical single-particle 
energies $\epsilon_j$ and the occupation probabilities $v_j^2$ are obtained from the RHB model. 
The same RHB model calculation that determines the entire 
($\beta, \gamma$)-deformation energy surface, when performed at zero deformation and with either the proton
or neutron number constrained to the desired odd number, but without 
blocking, gives the canonical single-particle energies and occupation probabilities
of the odd-fermion orbitals included in Tabs.~\ref{tab:spe} and
\ref{tab:vv}, respectively. Note that the exchange boson-fermion interaction in Eq.~(\ref{eq:bf}) 
takes into account the fact that the bosons are fermion pairs.

\begin{table}[hb!]
\caption{\label{tab:spe} Spherical single-particle energies for the
 $2d_{3/2}$, $2d_{5/2}$ and $1g_{7/2}$ orbitals (in MeV) relative to that of
 the $3s_{1/2}$ orbital, obtained in 
the RHB calculation for the odd-mass
nuclei considered in the present study.}
\begin{center}
\begin{tabular*}{\columnwidth}{p{2.0cm}p{2.0cm}p{2.0cm}p{2.0cm}}
\hline\hline
\textrm{} &
\textrm{$2d_{3/2}$}&
\textrm{$2d_{5/2}$}&
\textrm{$1g_{7/2}$} \\
\hline
$^{129}$Ba & 0.410 & 2.528 & 4.619 \\
$^{131}$Ba & 0.455 & 2.574 & 4.761 \\
$^{133}$Ba & 0.498 & 2.619 & 4.898 \\
$^{135}$Ba & 0.539 & 2.665 & 5.030 \\
$^{137}$Ba & 0.578 & 2.714 & 5.157 \\
$^{127}$Xe & 0.358 & 2.530 & 4.326 \\
$^{129}$Xe & 0.400 & 2.582 & 4.450 \\
$^{131}$Xe & 0.433 & 2.625 & 4.562 \\
$^{133}$Xe & 0.479 & 2.682 & 4.684 \\
$^{135}$Xe & 0.516 & 2.733 & 4.795 \\
$^{129}$La & -0.689 & -2.726 & -4.538 \\
$^{131}$La & -0.737 & -2.752 & -4.716 \\
$^{133}$La & -0.780 & -2.772 & -4.896 \\
$^{135}$La & -0.814 & -2.785 & -5.073 \\
$^{137}$La & -0.837 & -2.788 & -5.239 \\
$^{127}$Cs & -0.704 & -2.798 & -4.467 \\
$^{129}$Cs & -0.745 & -2.822 & -4.642 \\
$^{131}$Cs & -0.781 & -2.840 & -4.824 \\
$^{133}$Cs & -0.814 & -2.853 & -5.010 \\
$^{135}$Cs & -0.844 & -2.863 & -5.199 \\
\hline\hline
\end{tabular*}
\end{center}
\end{table}

\begin{table}[hb!]
\caption{\label{tab:vv} Occupation probabilities of the spherical single-particle
 orbitals obtained in the SCMF calculation for the odd-A isotopes.}
\begin{center}
\begin{tabular*}{\columnwidth}{p{1.33cm}p{1.33cm}p{1.33cm}p{1.33cm}p{1.33cm}p{1.33cm}}
\hline\hline
\textrm{} &
\textrm{$3s_{1/2}$}&
\textrm{$2d_{3/2}$}&
\textrm{$2d_{5/2}$}&
\textrm{$1g_{7/2}$}&
\textrm{$1h_{11/2}$} \\
\hline
$^{129}$Ba & 0.597 & 0.708 & 0.938 & 0.974 & 0.453 \\
$^{131}$Ba & 0.682 & 0.785 & 0.953 & 0.979 & 0.568 \\
$^{133}$Ba & 0.768 & 0.854 & 0.967 & 0.985 & 0.687 \\
$^{135}$Ba & 0.856 & 0.916 & 0.980 & 0.991 & 0.810 \\
$^{137}$Ba & 0.950 & 0.973 & 0.993 & 0.997 & 0.936 \\
$^{127}$Xe & 0.650 & 0.738 & 0.945 & 0.973 & 0.431 \\
$^{129}$Xe & 0.736 & 0.812 & 0.958 & 0.979 & 0.547 \\
$^{131}$Xe & 0.818 & 0.875 & 0.971 & 0.985 & 0.670 \\
$^{133}$Xe & 0.894 & 0.929 & 0.983 & 0.991 & 0.798 \\
$^{135}$Xe & 0.966 & 0.978 & 0.994 & 0.997 & 0.931 \\
$^{129}$La & 0.016 & 0.033 & 0.154 & 0.718 & 0.023 \\
$^{131}$La & 0.014 & 0.029 & 0.132 & 0.739 & 0.022 \\
$^{133}$La & 0.012 & 0.025 & 0.110 & 0.759 & 0.020 \\
$^{135}$La & 0.010 & 0.022 & 0.089 & 0.779 & 0.018 \\
$^{137}$La & 0.008 & 0.018 & 0.070 & 0.797 & 0.017 \\
$^{127}$Cs & 0.011 & 0.023 & 0.091 & 0.534 & 0.017 \\
$^{129}$Cs & 0.010 & 0.020 & 0.078 & 0.546 & 0.017 \\
$^{131}$Cs & 0.009 & 0.018 & 0.065 & 0.558 & 0.016 \\
$^{133}$Cs & 0.008 & 0.016 & 0.053 & 0.569 & 0.015 \\
$^{135}$Cs & 0.007 & 0.014 & 0.044 & 0.578 & 0.014 \\
\hline\hline
\end{tabular*}
\end{center}
\end{table}

Finally, the three strength constants of the boson-fermion interaction $\hat
H_{BF}$ ($\Gamma_0^{\pm}$, $\Lambda_0^{\pm}$ and $A_0^{\pm}$) are the only
phenomenological parameters and, for each nucleus, their values are adjusted to
reproduce a few lowest experimental states, separately for positive- and
negative-parity states \cite{nomura2016odd}.

In Tab.~\ref{tab:paraBFpos} we display the fitted strength parameters of
$\hat H_{BF}$ for the positive-parity states. 
The strength of the quadrupole dynamical term $\Gamma_0^+$ is
almost constant for each isotopic chain, except for the heaviest isotopes 
near the closed shell at $N=82$, whose structure differs significantly from the lighter ones. 
The strength parameter of the exchange term $\Lambda_0^+$ exhibits a
gradual variation (either increase or decrease) with neutron number. 
While in the phenomenological IBFM calculations \cite{cunningham1982b,dellagiacoma1988phdthesis} 
a $j$-independent monopole strength was used for all fermion orbitals,
in the present analysis, as in our previous study of Ref.~\cite{nomura2016odd},
the strength parameter of the monopole interaction is allowed to be $j$-dependent, $A_0\equiv A_j^{\prime}$ 
for positive-parity states. This is because the microscopic single-particle energies that we use in the 
present calculation are rather different from the empirical ones
employed in Refs.~\cite{cunningham1982b,dellagiacoma1988phdthesis}.  
In the case of $^{135}$Ba, for instance, the single-particle
orbital ${2d_{3/2}}$ is here calculated $\approx 0.5$ MeV above the ${3s_{1/2}}$ orbital.  
In the fully phenomenological model of  
Ref.~\cite{dellagiacoma1988phdthesis}, on the other hand, the ordering of the two orbitals
is reversed, that is, $\epsilon_{d_{3/2}}<\epsilon_{s_{1/2}}$. This is consistent 
with the empirical interpretation that the lowest and
second-lowest positive-parity states of $^{135}$Ba, with
$J^{\pi}={3/2}^+_1$ and ${1/2}^+_1$, are predominantly based on the 
$2d_{3/2}$ and $3s_{1/2}$ configurations, respectively. 
To reproduce the correct empirical level ordering of the lowest two
positive-parity states of $^{135}$Ba, here the monopole term is
adjusted specifically for the $2d_{3/2}$ orbital so that the
$J^{\pi}={3/2}^+_1$ state becomes the lowest positive-parity state. 
We have also verified that with an $j$-independent monopole strength 
the empirical low-lying positive-parity spectra of $^{135}$Ba 
cannot be reproduced.

For the negative-parity states (Tab.~\ref{tab:paraBFneg}), the three
strength parameters ($\Gamma_0^-$, $\Lambda_0^-$ and $A_0^-$) are either 
constant or change gradually with neutron number. 
Since the exchange term gives only a small contribution for the negative-parity
spectra of the odd-Z (La and Cs) isotopes, the exchange interaction
strength $\Lambda_0^-$ is set to zero.

\begin{table}[hb!]
\caption{\label{tab:paraBFpos} Parameters of the boson-fermion
 Hamiltonian $\hat H_{BF}$ for positive-parity states. All entries
 are in units of MeV. }
\begin{center}
\begin{tabular*}{\columnwidth}{p{1.14cm}p{1.14cm}p{1.14cm}p{1.14cm}p{1.14cm}p{1.14cm}p{1.14cm}}
\hline\hline
\textrm{} &
\textrm{$\Gamma_0^+$}&
\textrm{$\Lambda_0^+$}&
\textrm{$A_{1/2}^{\prime}$} &
\textrm{$A_{3/2}^{\prime}$}&
\textrm{$A_{5/2}^{\prime}$}&
\textrm{$A_{7/2}^{\prime}$} \\
\hline
$^{129}$Ba & 0.6 & 5.0 & -0.21 & {} & {} & -0.88 \\
$^{131}$Ba & 0.6 & 3.5 & -0.09 & {} & {} & {} \\
$^{133}$Ba & 0.6 & 3.5 & {} & -0.05 & {} & {} \\
$^{135}$Ba & 0.6 & 2.0 & {} & -0.55 & {} & {} \\
$^{137}$Ba & 2.0 & 1.0 & {} & -1.3 & {} & {} \\
$^{127}$Xe & 0.6 & 4.0 & -0.28 & {} & {} & -0.92 \\
$^{129}$Xe & 0.6 & 2.5 & -0.12 & {} & {} & -1.01 \\
$^{131}$Xe & 0.4 & 2.0 & {} & -0.27 & {} & {} \\
$^{133}$Xe & 1.5 & 1.0 & {} & -0.95 & {} & {} \\
$^{135}$Xe & 2.0 & 1.0 & {} & -1.35 & {} & {} \\
$^{129}$La & 0.2 & 1.15 & {} & {} & -1.20 & {} \\
$^{131}$La & 0.2 & 1.25 & {} & {} & -1.25 & {} \\
$^{133}$La & 0.2 & 1.5 & {} & {} & -0.82 & {} \\
$^{135}$La & 0.2 & 2.2 & {} & {} & -1.11 & {} \\
$^{137}$La & 0.01 & 3.0 & {} & {} & -1.5 & {} \\
$^{127}$Cs & 0.4 & 2.2 & {} & {} & -1.5 & {} \\
$^{129}$Cs & 0.4 & 1.85 & {} & {} & -1.5 & {} \\
$^{131}$Cs & 0.4 & 1.0 & {} & {} & -0.7 & {} \\
$^{133}$Cs & 0.2 & 1.3 & {} & {} & -1.05 & {} \\
$^{135}$Cs & 0.2 & 1.3 & {} & {} & -1.35 & {} \\
\hline\hline
\end{tabular*}
\end{center}
\end{table}

\begin{table}[hb!]
\caption{\label{tab:paraBFneg} Same as in the caption to Tab.~\ref{tab:paraBFpos}, but for
negative-parity states. }
\begin{center}
\begin{tabular*}{\columnwidth}{p{2.0cm}p{2.0cm}p{2.0cm}p{2.0cm}}
\hline\hline
\textrm{} &
\textrm{$\Gamma_0^-$}&
\textrm{$\Lambda_0^-$}&
\textrm{$A_{11/2}^{\prime}$} \\
\hline
$^{129}$Ba & 0.6 & 2.1 & -0.15 \\
$^{131}$Ba & 0.6 & 2.1 & -0.23 \\
$^{133}$Ba & 0.6 & 0.9 & 0.0 \\
$^{135}$Ba & 0.6 & 1.0 & -0.9 \\
$^{137}$Ba & 0.4 & 5.0 & -0.6 \\
$^{127}$Xe & 0.6 & 2.0 & -0.2 \\
$^{129}$Xe & 0.6 & 1.7 & -0.13 \\
$^{131}$Xe & 0.6 & 1.6 & -0.10 \\
$^{133}$Xe & 0.4 & 1.0 & -0.20 \\
$^{135}$Xe & 0.45 & 0.0 & 0.0 \\
$^{129}$La & 0.1 & 0.0 & -0.3 \\
$^{131}$La & 0.1 & 0.0 & -0.28 \\
$^{133}$La & 0.1 & 0.0 & 0.0 \\
$^{135}$La & 0.1 & 0.0 & 0.0 \\
$^{137}$La & 0.1 & 10.0 & -0.2 \\
$^{127}$Cs & 0.1 & 0.0 & -0.07 \\
$^{129}$Cs & 0.1 & 0.0 & -0.11 \\
$^{131}$Cs & 0.1 & 0.0 & -0.05 \\
$^{133}$Cs & 0.6 & 0.0 & -0.20 \\
\hline\hline
\end{tabular*}
\end{center}
\end{table}

The resulting IBFM Hamiltonian $\hat H$ is diagonalized in the
spherical basis $|j, L, \alpha, J\rangle$ using the code PBOS
\cite{PBOS}, where $\alpha=(n_d,\nu,n_{\Delta})$ is a 
generic notation for the boson quantum numbers in the U(5) symmetry limit \cite{IBM},
that distinguish states with the same angular momentum of the boson system $L$. 
$J$ is the total angular momentum of the coupled boson-fermion system,
and satisfies the condition $|L-j|\leq J\leq L+j$.

By using the corresponding eigenfunctions, electromagnetic decay properties,
such as E2 and M1 transition rates, and spectroscopic quadrupole
and magnetic moments, are calculated for the odd-mass systems. 
The E2 operator contains the boson and fermion terms $\hat T^{(E2)}=\hat T^{(E2)}_B+\hat T^{(E2)}_F$. 
The expression for the IBM boson E2 operator:  
\begin{eqnarray}
 \hat T^{(E2)}_B = e_B(s^{\dagger}\tilde d+d^{\dagger}\tilde s +
\chi^{\prime}[d^{\dagger}\times\tilde d]^{(2)})
\end{eqnarray}
where $e_B$ is the boson effective charge and $\chi^{\prime}$ is a parameter. 
The fermion E2 operator used in the present calculation reads:  
\begin{eqnarray}
 \hat T^{(E2)}=
-e_F\sum_{jj^{\prime}}
\frac{1}{\sqrt{5}}
\gamma_{jj^{\prime}}
[a^{\dagger}\times\tilde a_{j^{\prime}}]^{(2)},
\end{eqnarray}
with the fermion effective charge $e_F$. As in many phenomenological
studies and also in our previous articles on shape-phase transitions in odd-A nuclei \cite{nomura2016odd,nomura2016qpt},
the effective charge $e_B$ is determined by the experimental value of $B(E2; 2^+_1\rightarrow 0^+_1)$ in each 
even-even core nucleus. The parameter $\chi^{\prime}$ is adjusted to reproduce the
experimental spectroscopic quadrupole moment of the $2^+_1$ state
(denoted as $Q_{2^+_1}$) of $^{136}$Ba, and is fixed to the value
$\chi^{\prime}=0.35$ for all nuclei considered in the present study. 
Finally, the value of the fermion effective charges $e_F$ are adjusted to 
the experimental values of $Q_{{5/2}^+_1}$ of $^{137}$La and
$Q_{{11/2}^-_1}$ of $^{137}$Ba. The corresponding proton $e_p=0.250\,e$b
and neutron $e_n=0.125\,e$b effective charges are 
used for the odd-Z nuclei and odd-N nuclei, respectively. These values are consistent 
with standard IBFM calculations performed in this and other mass regions
\cite{IBFM-Book,dellagiacoma1988phdthesis,yoshida1989,abumusleh2014}, as well as with the
microscopic analysis of the IBFM \cite{scholten1982}. 
The M1 operator is given by
\begin{eqnarray}
 \hat T^{(M1)}=\sqrt{\frac{3}{4\pi}}(\hat T^{(M1)}_B + \hat T^{(M1)}_F)
\end{eqnarray}
where $\hat T^{(M1)}_B=g_B\hat L$ is the boson M1 operator, and  
the fermion operator $\hat T^{(M1)}_F$ \cite{scholten1985}: 
\begin{eqnarray}
 \hat T^{(M1)}_F=-\sum_{jj^{\prime}} g_{jj^{\prime}}\sqrt{\frac{j(j+1)(2j+1)}{3}}[a^{\dagger}_j\times\tilde a_{j^{\prime}}]^{(1)},
\end{eqnarray}
with
\begin{eqnarray}
 g_{jj^{\prime}}
=\left\{\begin{array}{ll}
\frac{(2j-1)g_l + g_s}{2j} & (j=j^{\prime}=l+\frac{1}{2}) \\
  \frac{(2j+3)g_l - g_s}{2(j+1)} & (j=j^{\prime}=l-\frac{1}{2}) \\
	 (g_l-g_s)\sqrt{\frac{2l(l+1)}{j(j+1)(2j+1)(2l+1)}} &
	  (j^{\prime}=j-1; l=l^{\prime}) \\
	\end{array} \right.,
\end{eqnarray}
and $l$ is the orbital angular momentum of the single-particle state. 
The value of the boson $g$-factor is $g_B=\mu_{2^+_1}/2$, where
$\mu_{2^+_1}$ is the magnetic moment of the state $2^+_1$ of the
even-even nucleus, and the corresponding experimental value is used for 
this quantity. 
For the fermion $g$-factors: $g_l=1.0$ $\mu^2_N$ for the odd proton, 
and $g_l=0$ for the odd neutron, and 
free values of $g_s$ are quenched by 30 \% as used, for instance, 
in Refs.~\cite{scholten1982,nomura2016odd}.

Summarizing this section, we note that the IBFM
Hamiltonian (\ref{eq:ham}) in the present implementation  
contains altogether twenty-two parameters. While the parameters of the 
boson and fermion Hamiltonians are determined by the microscopic 
self-consistent mean-field calculation, nine parameters: 
$\Gamma_0^{\pm}$, $\Lambda_0^{\pm}$ and $A_j^{\prime}$ for five
orbitals, are specifically adjusted to experimental low-energy excitation spectra.
In addition, the four parameters $e_B$, $\chi^{\prime}$,
$e_p$ and $e_n$ of the E2 operator, are adjusted 
to reproduce specific E2 data.


\section{Signatures of shape phase transitions in the odd-A $\gamma$-soft nuclei \label{sec:results}}


\begin{figure}[htb!]
\begin{center}
\includegraphics[width=\linewidth]{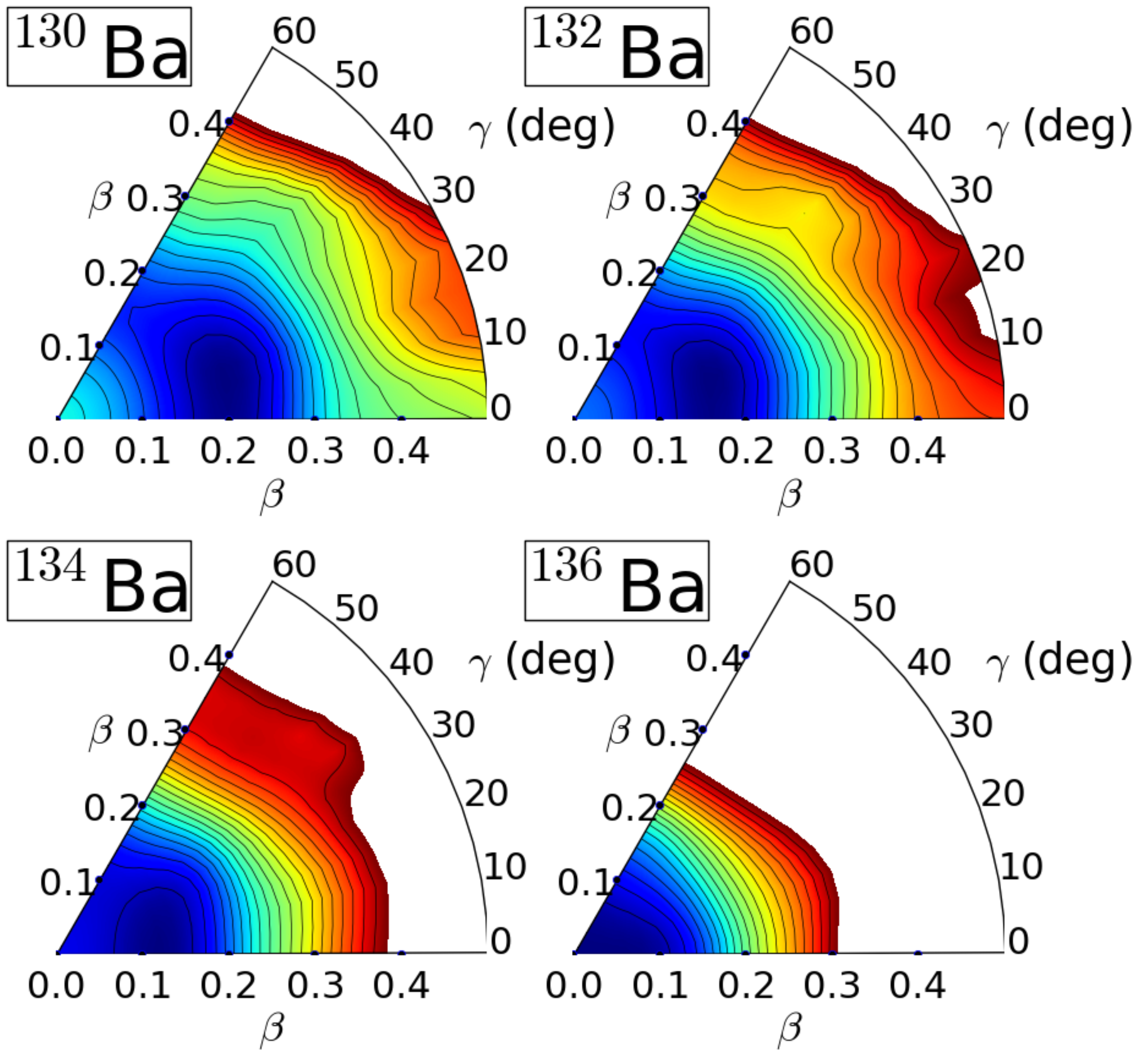}
\caption{(Color online) Self-consistent RHB triaxial quadrupole
binding energy maps of the even-even $^{130-136}$Ba isotopes
in the $\beta - \gamma$ plane ($0\le \gamma\le 60^{\circ}$).
For each nucleus the energy surface is normalized with respect to
the binding energy of the absolute minimum, and is plotted up to 10 MeV
excitation energy with 0.2 MeV difference between neighboring
 contours.}
\label{fig:evenba-pes}
\end{center}
\end{figure}

\begin{figure}[htb!]
\begin{center}
\includegraphics[width=\linewidth]{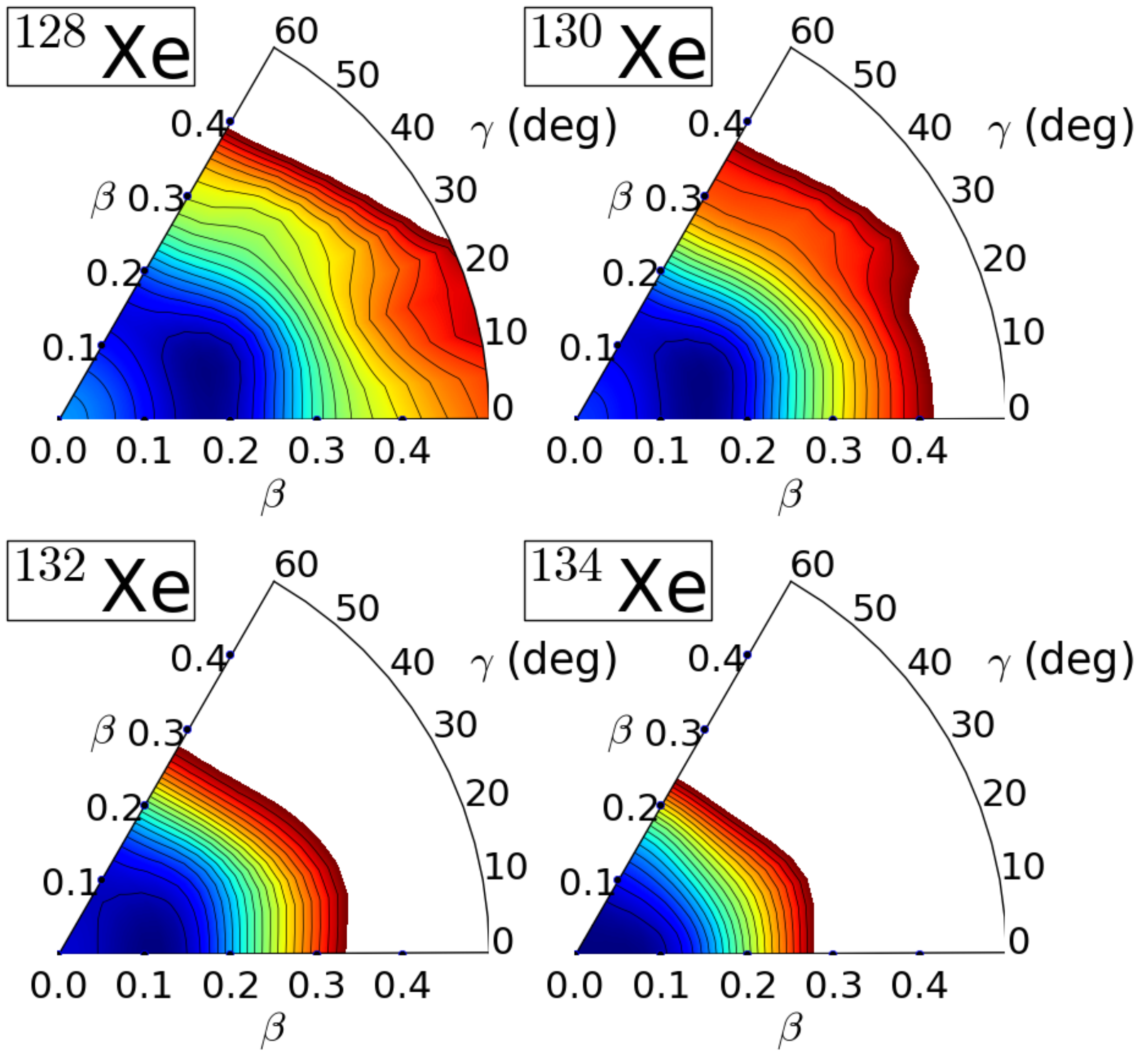}
\caption{(Color online) Same as in the caption to Fig.~\ref{fig:evenxe-pes}, but for 
 $^{128-134}$Xe.}
\label{fig:evenxe-pes}
\end{center}
\end{figure}

\subsection{Deformation energy surface}

As explained in the previous section, the deformation energy surfaces 
for a set of even-even Ba and Xe isotopes that determine the parameters of the 
IBM Hamiltonian, are calculated as functions
of the polar deformation parameters $\beta$ and $\gamma$ \cite{BM}, using 
the constrained relativistic Hartree-Bogoliubov method based on the
functional DD-PC1 \cite{DDPC1} and a separable pairing force of finite
range \cite{tian2009}. 
A triaxial binding energy map as a function of quadrupole shape variables is
obtained by imposing constraints on both the axial and triaxial mass
quadrupole moments. 
In Figs.~\ref{fig:evenba-pes} and \ref{fig:evenxe-pes}, the 
energy surfaces for the even-even core nuclei $^{130-136}$Ba 
and $^{128-134}$Xe, respectively, are displayed in the $\beta-\gamma$
plane ($0^{\circ}\leq\gamma\leq 60^{\circ}$). 
We note that the energy surfaces for the $^{128}$Ba and $^{126}$Xe nuclei are
nearly identical to those of their adjacent nuclei $^{130}$Ba and $^{128}$Xe,
respectively, and thus are not included in the figures. 

At the self-consistent mean-field level the RHB energy surfaces display a 
gradual transition of equilibrium shapes as a function of the (valence)
neutron number. One notices 
that the RHB energy surfaces for the Ba and Xe isotopes are very 
similar and, for this reason, we discuss only the results for the Ba
isotopes. 
As shown in Fig.~\ref{fig:evenba-pes}, the shape is noticeably soft in
$\gamma$ deformation for $^{130,132}$Ba with a very shallow triaxial
minimum in the interval $\gamma=10^{\circ}-20^{\circ}$. 
As the number of valence nucleons (neutron holes) decreases for
$^{134}$Ba, the potential appears to become almost completely flat in
$\gamma$ direction, which is a typical feature of transitional nuclei. 
$^{136}$Ba displays a nearly spherical shape with a
minimum at $\beta\approx 0.1$, reflecting the $N=82$ neutron shell closure. 
It is interesting that the equilibrium
shapes for the Ba nuclei display no significant change in the axial
deformation $\beta$ as a function of the neutron number. We also 
note that the RHB energy surfaces for the Xe isotopes appear to be somewhat softer
in $\gamma$ when compared to the corresponding Ba neighbors.

In the present analysis we are particularly interested in transitional nuclei. 
$^{134}$Ba is located between the
nearly spherical shapes close to $N=82$ and the $\gamma$-soft shapes 
of lighter isotopes. 
This nucleus was analyzed as the first empirical realization \cite{casten2000}
of the critical point of second-order QPT 
between spherical and $\gamma$-soft shapes, described by the E(5) symmetry
\cite{iachello2000}. This symmetry corresponds to the five-dimensional collective
Hamiltonian (the intrinsic variables $\beta$ and $\gamma$ and the three Euler angles), 
with an infinite square-well potential in
the axial deformation $\beta$, and independent of $\gamma$ \cite{iachello2000}. 
One notices that the microscopic deformation energy surface of $^{134}$Ba
in the present calculation is closest to the E(5)-like
potential: it is flat-bottomed for small values of the axial deformation 
$\beta < 0.2$, and almost completely flat in the $\gamma$ direction. 
A similar shape is predicted for $^{132}$Xe.

\subsection{Low-energy excitation spectra}

\begin{figure}[htb!]
\begin{center}
\includegraphics[width=\linewidth]{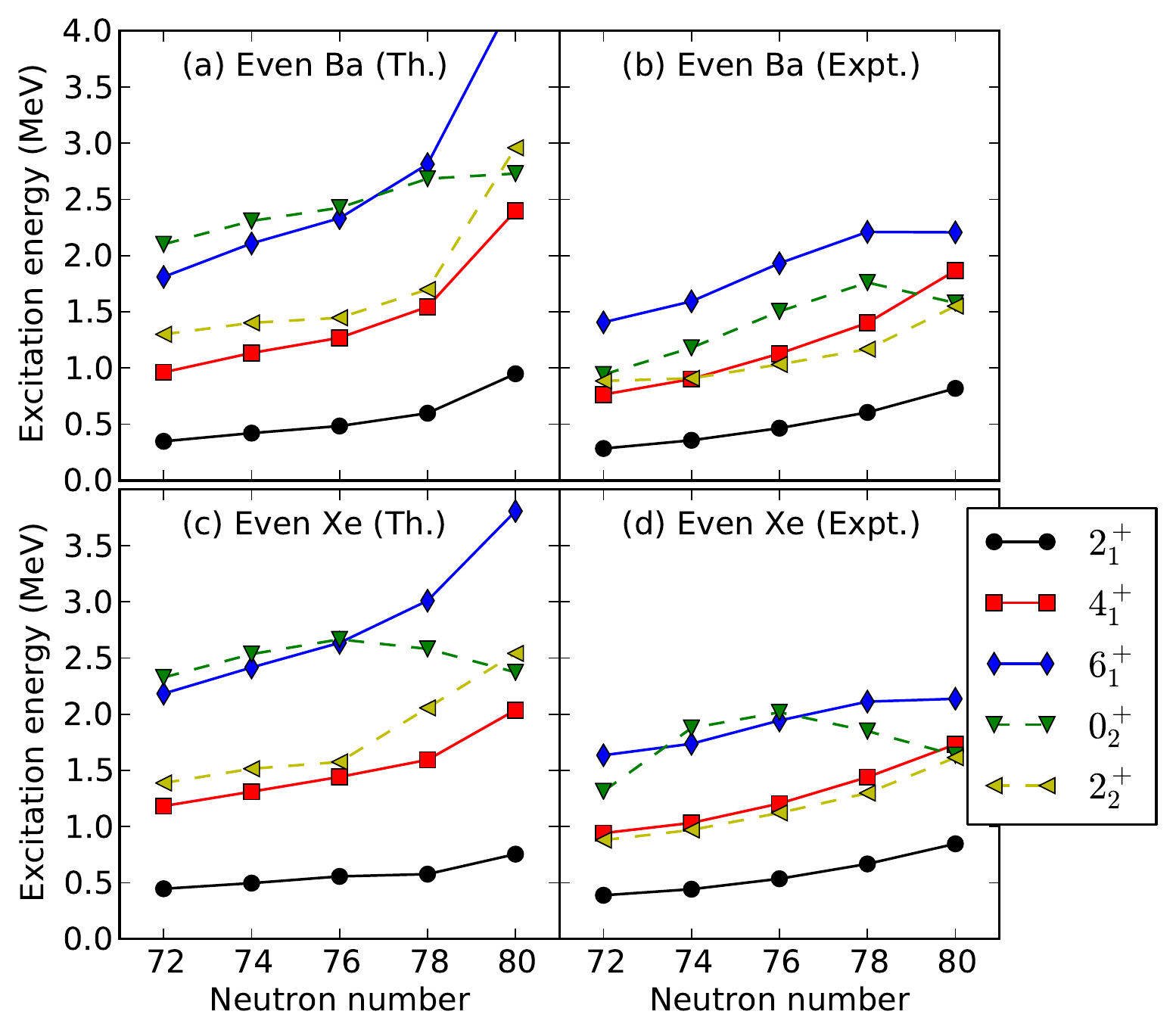}
\caption{(Color online) Evolution of low-lying collective states in $^{128-136}$Ba
and  $^{126-134}$Xe, as functions of the neutron
 number. Theoretical excitation spectra in the left column are compared 
 to data (panels on the right) from Ref.~\cite{data}.}
\label{fig:even-level}
\end{center}
\end{figure}

\begin{figure}[htb!]
\begin{center}
\includegraphics[width=\linewidth]{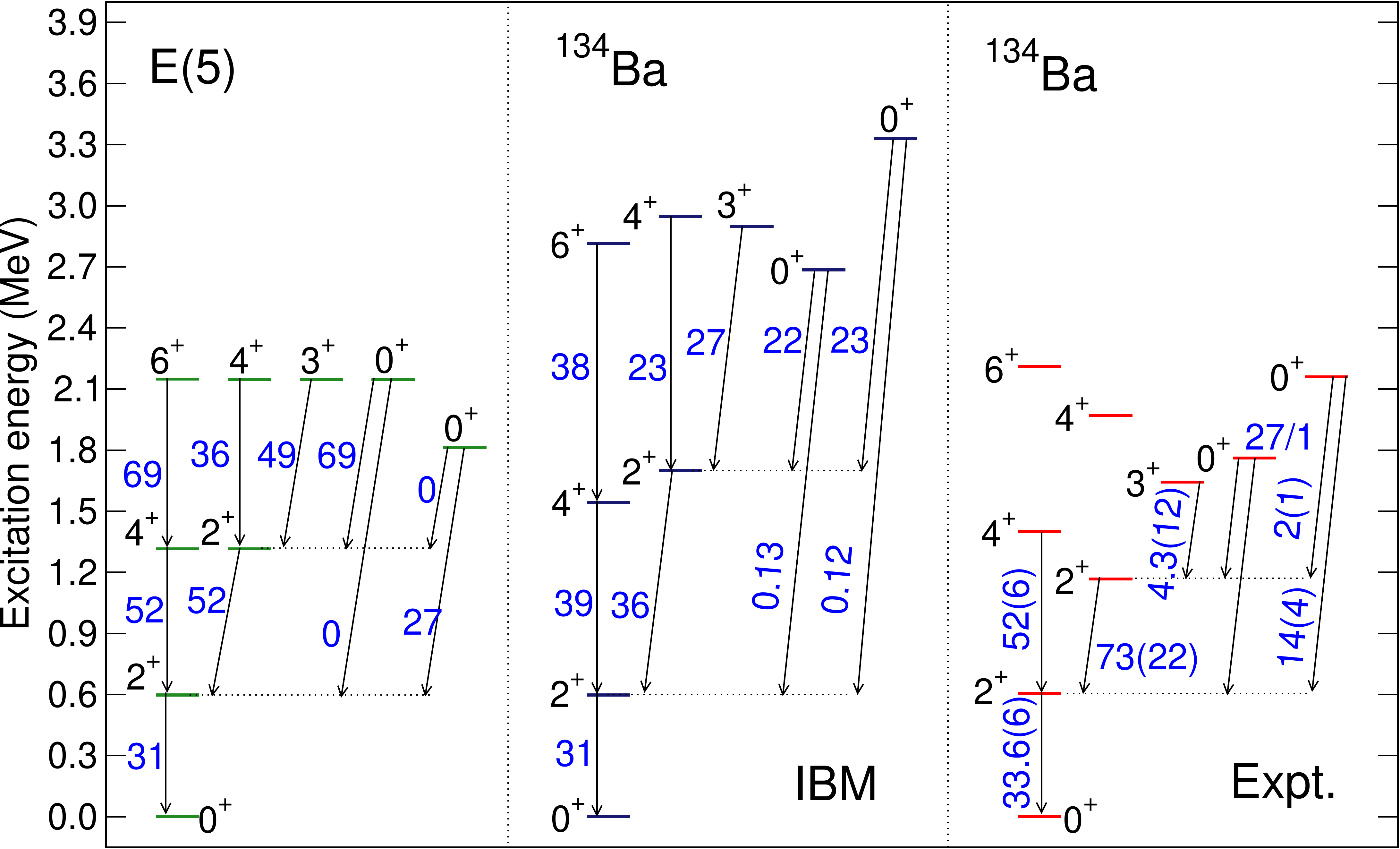}
\caption{(Color online) The low-energy spectrum of $^{134}$Ba 
calculated with the microscopic IBM, in comparison to 
available data \cite{data,casten2001}. 
The excitation spectrum that corresponds to the E(5) symmetry limit is also displayed, with
 the excitation energy of the state $2^{+}_1$ and the $B(E2;
 2^+_1\rightarrow 0^+_1)$ transition strength normalized to the values obtained 
 in the IBM calculation.}
\label{fig:ba134-E5}
\end{center}
\end{figure}


A QPT is characterized by a significant
variation of order parameters as functions of the physical control parameter. 
While the analysis of potential energy surfaces provides an
approximate indication of QPT at the mean-field level, the
intrinsic deformation parameters are not observables and a quantitative
analysis of the nuclear phase transitions must, therefore, extend beyond the simple Landau
approach to include a direct calculation of observables that can be
interpreted as quantum order parameters. 

To illustrate the level of accuracy with which the boson-core Hamiltonian, with parameters 
determined by mapping the microscopic energy surface onto the expectation
value of the IBM Hamiltonian, describes spectroscopic properties of
even-even systems, we begin by comparing in Fig.~\ref{fig:even-level} the computed
excitation spectra for the low-lying states of the even-even $^{128-136}$Ba
and $^{126-134}$Xe isotopes to available data \cite{data}. 
Evidently the model calculation reproduces the empirical
systematics of low-lying excitation spectra. 
In particular, the $\gamma$-softness of the effective nuclear 
potential is characterized by close-lying $4^+_1$ and $2^+_2$ levels. 
Both experimentally and in model calculations, this level structure is observed from
$N=72$ up to 78.  At $N=80$ the energy spacings correspond to vibrational spectra, as identified
by the multiplets of levels $(4^+_1, 2^+_2, 0^+_2)$. 
Overall, the theoretical excitation spectra are more stretched than the
experimental ones, especially at $N=80$. 
This could be attributed to the limited IBM configuration space
consisting only of the valence nucleon pairs outside closed shells.

$^{134}$Ba is considered an excellent example of empirical realization of the E(5)
critical-point symmetry \cite{casten2000}. 
In Fig.~\ref{fig:ba134-E5} we compare the calculated energy
spectrum of this nucleus with the experimental low-energy levels, as well as
with the spectrum corresponding to the E(5) symmetry limit. 
In comparison to the experimental levels, the present calculation
generally predicts higher excitation energies, but exhibits several
features that correspond to the E(5) symmetry, including the
close-lying ($4^+_1,2^+_2$) and ($6^+_1,4^+_2,3^+_1,0^+_2$) levels, as well as the
selection rule for E2 transitions from the $0^+_2$ to the $2^+_{1,2}$ states. 

\begin{figure}[htb!]
\begin{center}
\includegraphics[width=\linewidth]{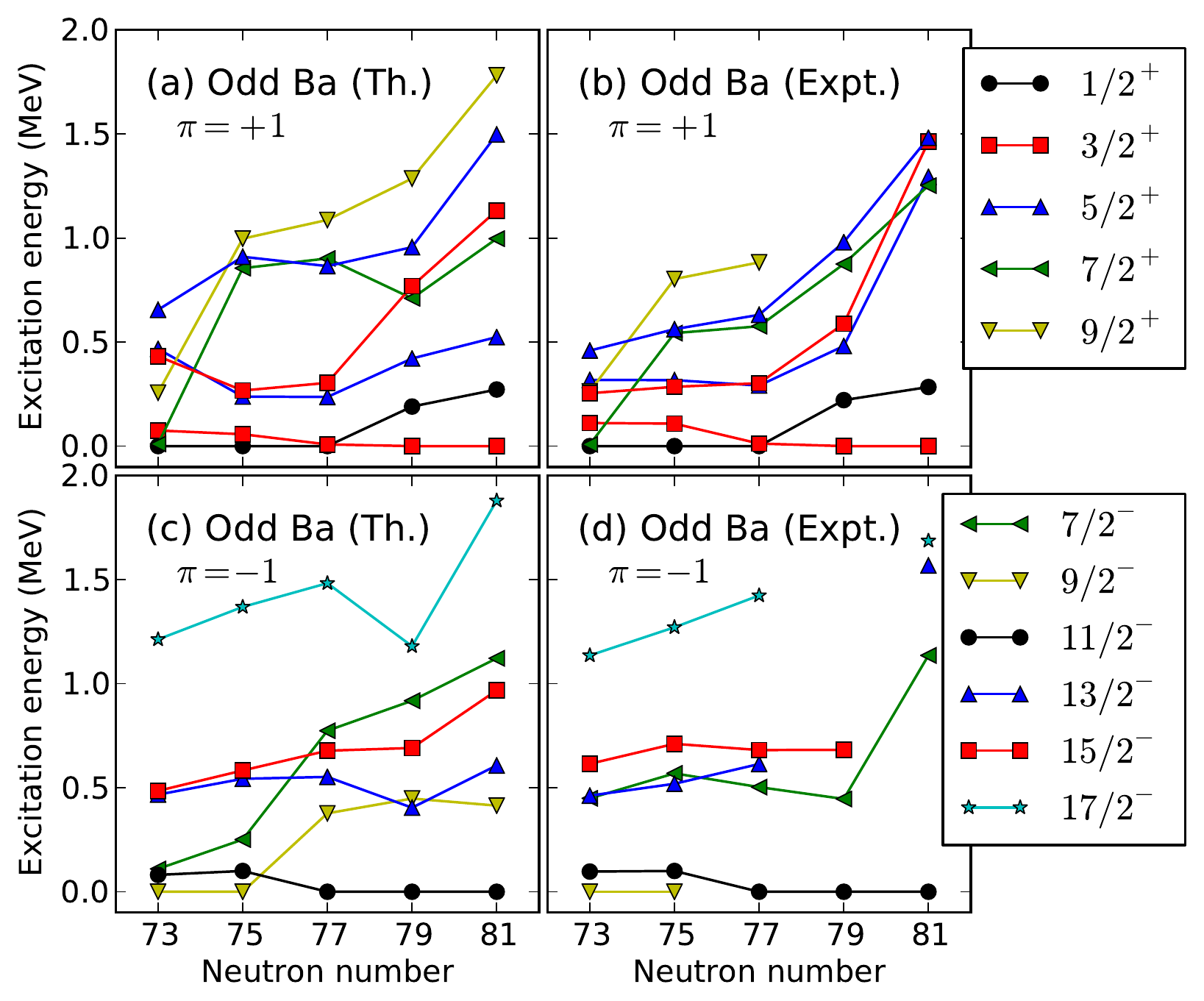}
\caption{(Color online) Evolution of the low-lying positive- and
 negative-parity states in the odd-A isotopes $^{129-137}$Ba
 as functions of the neutron
 number. The experimental levels are from Ref.~\cite{data}.}
\label{fig:oddba-level}
\end{center}
\end{figure}

\begin{figure}[htb!]
\begin{center}
\includegraphics[width=\linewidth]{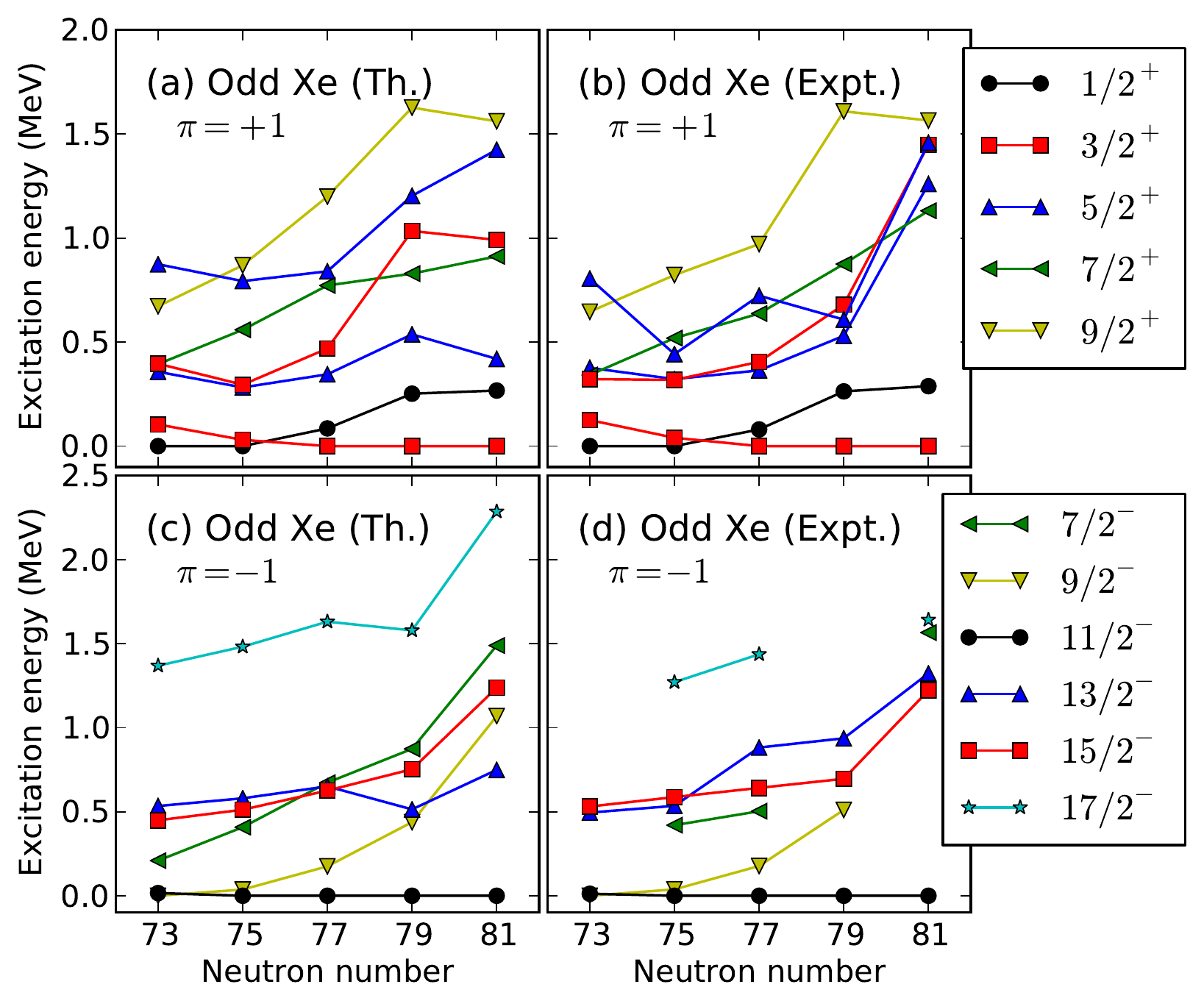}
\caption{(Color online) Same as in the caption to Fig.~\ref{fig:oddba-level}, but for the
 isotopes $^{127-135}$Xe.}
\label{fig:oddxe-level}
\end{center}
\end{figure}

\begin{figure}[htb!]
\begin{center}
\includegraphics[width=\linewidth]{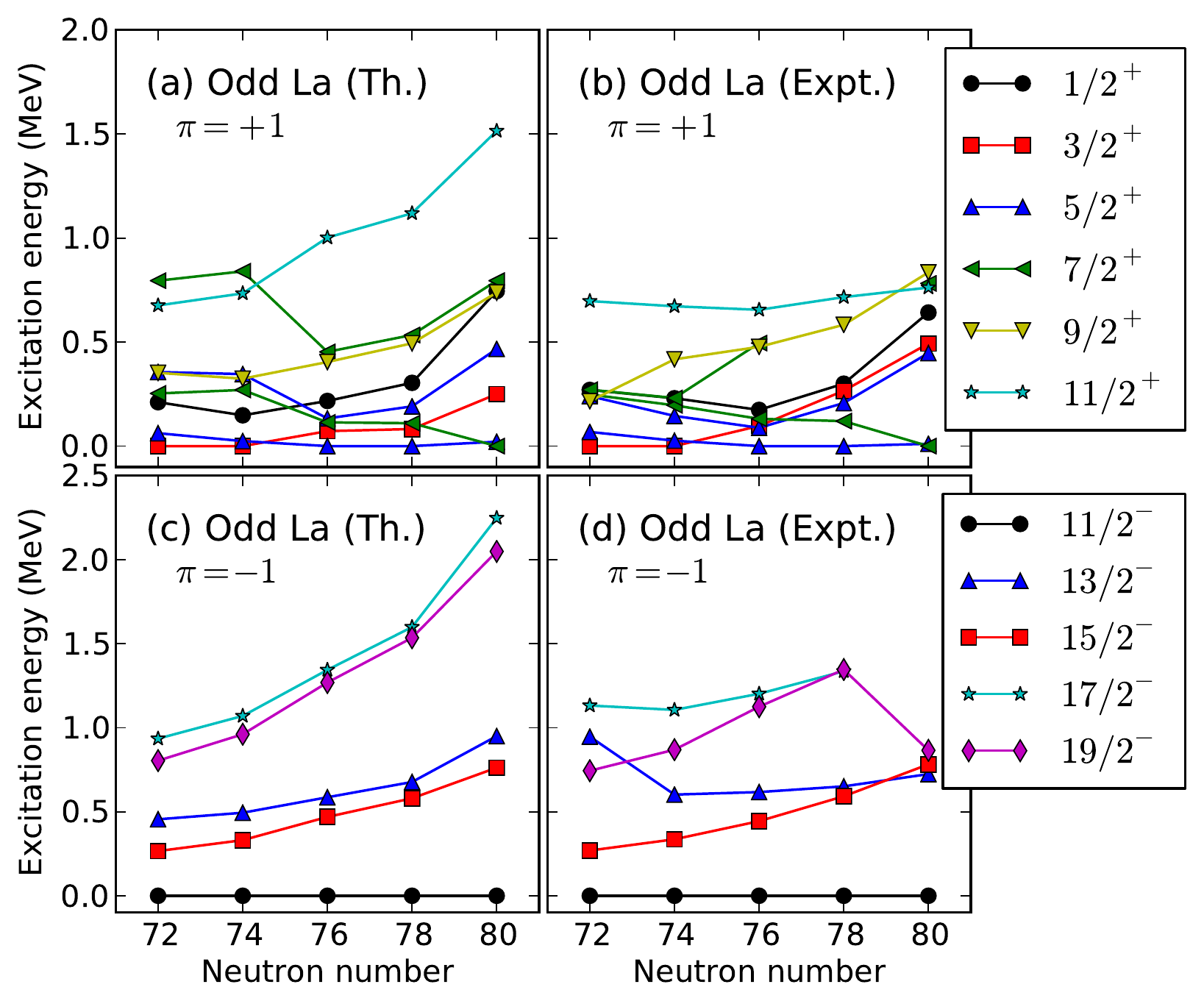}
\caption{(Color online) Same as in the caption to Fig.~\ref{fig:oddba-level}, but for
 $^{129-137}$La.}
\label{fig:oddla-level}
\end{center}
\end{figure}

\begin{figure}[htb!]
\begin{center}
\includegraphics[width=\linewidth]{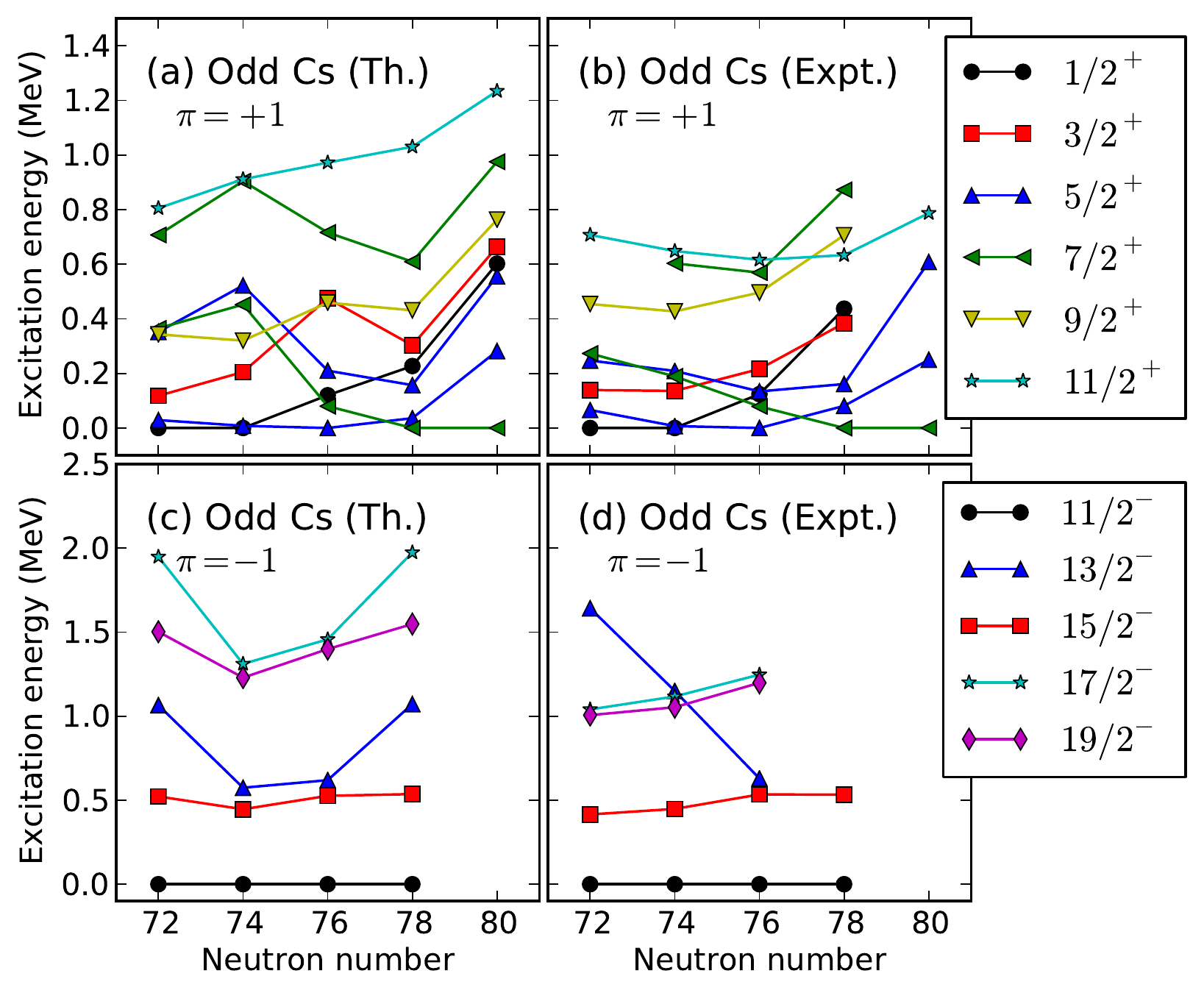}
\caption{(Color online) Same as in the caption to Fig.~\ref{fig:oddba-level}, but for  $^{127-135}$Cs.}
\label{fig:oddcs-level}
\end{center}
\end{figure}

In the following we focus the analysis on the results for odd-A systems. 
Figures~\ref{fig:oddba-level}-\ref{fig:oddcs-level} display the
calculated low-energy positive ($\pi=+1$) and negative-parity
($\pi=-1$) levels of the odd-A isotopes $^{129-137}$Ba, $^{127-135}$Xe,
$^{129-137}$La and $^{127-135}$Cs, respectively, as functions of
the neutron number, in comparison with the experimental excitation spectra \cite{data}. 
We note a remarkable agreement between theory and experiment for both 
$\pi=+1$ and $\pi=-1$ states in all four isotopic chains.

A specific signature of QPT in odd-A nuclei is the change of the ground-state 
spin at a nucleon number that corresponds to the phase transition. 
For the odd-A Ba isotopes show in Fig.~\ref{fig:oddba-level}, for instance,
the spin of the lowest positive-parity state changes from $J^{\pi}={1/2}^+$ to
${3/2}^+$ at $N=79$, while the change of the lowest
negative-parity state from $J^{\pi}={9/2}^-$ to ${11/2}^-$ is observed at $N=77$. 
This result is in agreement with the assumption that the QPT  
in the even-even Ba isotopes occurs at $N=78$, that is, for $^{134}$Ba. It also 
illustrates the difficulty in locating the point of shape-phase transition 
when the physical control parameter (neutron number in this case) 
is not continuous. One also notices in Figs.~\ref{fig:oddba-level} (a) and (b) that,
compared to the other odd-A Ba isotopes considered, the
${7/2}^+_1$ and ${9/2}^+_1$ states at $N=73$ are noticeably low in
energy, almost degenerate with the ${1/2}^+_1$ ground state. 
Empirically, it has been suggested that these states predominantly correspond to 
the $1g_{7/2}$ configuration \cite{data,cunningham1982b}, reflecting the fact 
that the $1g_{7/2}$ single-particle orbital is particularly low at $N=73$, and 
close in energy to the $3s_{1/2}$ and $2d_{3/2}$ orbitals. 
In our analysis, the calculated wave functions of the ${7/2}^+_1$ and
${9/2}^+_1$ states are almost pure (94 and 96 \%, respectively) $1g_{7/2}$
configurations, which conforms to the empirical interpretation of these
states. Figure~\ref{fig:oddxe-level} displays a similar pattern for the odd-A Xe isotopes, 
except that in this case the change in spin of the lowest positive and negative parity 
states occurs already at $N=77$ and $N=75$, respectively.

In the odd-Z systems $^{129-137}$La (Fig.~\ref{fig:oddla-level}) and
 $^{127-135}$Cs (Fig.~\ref{fig:oddcs-level}), on the one hand we notice the crossings between low-energy 
 positive-parity levels in the transitional region between $N=76$ and $N=78$. 
On the other hand, the negative-parity states of both odd-A La and Cs
isotopes exhibit essentially the same level structure throughout the
isotopic chains, that is, the band
built on the ${11/2}^-$ state that follows the $\Delta J=2$ systematics
of the weak coupling limit.

\subsection{Detailed level schemes of selected odd-A nuclei}

The details of the IBFM results are illustrated for one odd-A nucleus of each isotopic chain:
$^{135}$Ba, $^{129}$Xe, $^{133}$La, and $^{131}$Cs. These specific nuclei are 
close to the shape-phase transition point, their low-energy level sequences are experimentally 
well established, and there is sufficient data to compare
with model results, especially for the $E2$ and $M1$ transitions, as well as spectroscopic moments. 
Note that the calculated levels are classified into bands according
to the dominant E2 decay branch. 


\begin{figure}[htb!]
\begin{center}
\includegraphics[width=\linewidth]{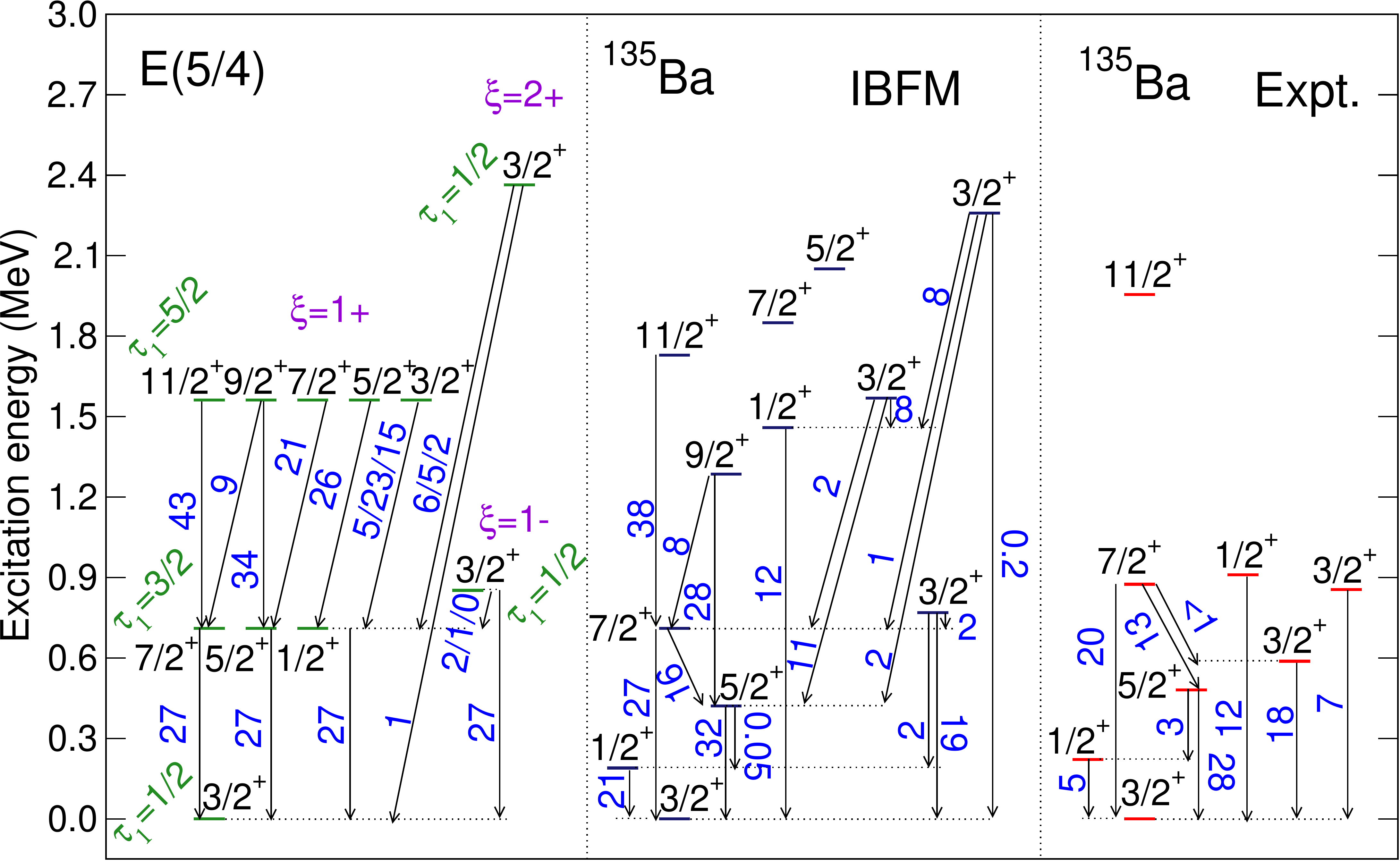}
\caption{(Color online) The calculated low-energy positive-parity spectrum of 
 $^{135}$Ba (IBFM), compared to the corresponding 
 experimental \cite{fetea2006} and E(5/4) excitation spectra. 
The quantum numbers of the E(5/4) model are also shown. Note that the energy 
 of the $(\xi=1+,\tau_1={3/2})$ E(5/4) multiplet, i.e., (${7/2}^+$, ${5/2}^+$, ${1/2}^+$),  
 is normalized to that of the ${7/2}^+_1$ IBFM state. $B(E2)$ values are 
 given in Weisskopf units, and the $B(E2)$ from the
 $(\xi=1+,\tau_1={3/2})$ E(5/4) multiplet is normalized to the $B(E2;
 {7/2}^+_1\rightarrow{3/2}^+_1)$ value obtained in the present IBFM calculation.
 The triplets of $B(E2)$ values in the E(5/4) model spectrum refer to transitions to the 
 ${7/2}^+_1$, ${5/2}^+_1$, and ${1/2}^+_1$ states, respectively.}
\label{fig:ba135}
\end{center}
\end{figure}

$^{135}$Ba is of particular interest in the present analysis, since the corresponding
even-even core $^{134}$Ba can be, to a good approximation, characterized by the E(5)
critical-point symmetry of the second-order QPT. 
In Ref.~\cite{iachello2005} the E(5/4) model of critical-point symmetry 
for odd-mass systems was developed, based on the concept of dynamical supersymmetry. 
The E(5/4) model describes the coupling of an unpaired $j={3/2}$ nucleon to  
the even-even boson core with E(5) symmetry. In fact, 
the first test of the E(5/4) Bose-Fermi symmetry \cite{fetea2006} considered 
the low-energy spectrum of $^{135}$Ba in terms of the neutron $2d_{3/2}$ orbital 
coupled to the E(5) boson core $^{134}$Ba.
In Fig.~\ref{fig:ba135} we compare the IBFM low-energy positive-parity
spectrum of $^{135}$Ba and the corresponding $B(E2)$ values with the 
predictions of the E(5/4) model, as well as with the
experimental excitation spectrum \cite{fetea2006}. Evidently 
the E(5/4) spectrum is more regular, that is, it displays degenerate multiplets of 
excited states, when compared to both the present IBFM and experimental energy spectra.  
Moreover, the E2 branching ratios of the E(5/4) model, e.g., from the excited ${3/2}^+$
states, differ from those obtained in the present calculation. 
This is not surprising because E(5/4) presents a simple scheme that 
takes into account only a single neutron valence orbit $2d_{3/2}$. 
In the phenomenological IBFM calculation that was carried out in
Ref.~\cite{fetea2006}, the wave functions of the ${1/2}^+_1$ and
${1/2}^+_2$ states were found to be mainly composed of the $3s_{1/2}$ and $2d_{3/2}$
configurations, respectively, and it was thus suggested that the ${1/2}^+_1$
state in the first excited E(5/4) multiplet should be compared with the
experimental ${1/2}^+_2$ state. 
Similar results are also obtained in the present calculation, as the
$3s_{1/2}$ and $2d_{3/2}$ configurations account for 58\% and 78\% 
of the wave functions of the ${1/2}^+_1$ and ${1/2}^+_2$ states,
respectively.

The present IBFM results reproduce the experimental 
excitation spectrum rather well, except for
the fact that several non-yrast states, such as ${1/2}^+_2$, are calculated at higher excitation energies. 
In Tab.~\ref{tab:ba135} we also compare in detail the calculated $B(E2)$ and
$B(M1)$ transition strengths, as well as the spectroscopic quadrupole
($Q_J$) and magnetic ($\mu_J$) moments, with available data \cite{data}. 
Considering the complexity of the level scheme and the large valence neutron space, 
a relatively good agreement is obtained between the calculated and experimental 
electromagnetic properties.

\begin{table}[htb]
\caption{\label{tab:ba135}%
Comparison between the theoretical and experimental $B(E2)$ and $B(M1)$
values, and spectroscopic
 quadrupole and magnetic moments in $^{135}$Ba. The data
are from Ref.~\cite{data}.}
\begin{center}
\begin{tabular}{p{2.0cm}cccc}
\hline\hline
\multirow{2}{*}{} & \multicolumn{2}{c}{$B(E2)$ (W.u.)} &
 \multicolumn{2}{c}{$B(M1)$ (W.u.)} \\
\cline{2-3} 
\cline{4-5}
          & Th.         & Expt.    & Th.         & Expt.     \\
\hline
${1/2}^+_1\rightarrow {3/2}^+_1$ & 21 & 4.6(2) & 0.0014 & 0.0025(11) \\
${1/2}^+_2\rightarrow {3/2}^+_1$ & 12 & 11.7(10) & - & - \\
${3/2}^+_2\rightarrow {3/2}^+_1$ & 2.0 & 18.0(10) & - & - \\
${3/2}^+_3\rightarrow {3/2}^+_1$ & 4.6 & 7.0(10) & - & - \\
${5/2}^+_1\rightarrow {1/2}^+_1$ & 0.05 & 2.6(5) & - & - \\
${5/2}^+_1\rightarrow {3/2}^+_1$ & 32 & 28.3(10) & 0.0012 & 0.0042(20) \\
${7/2}^+_1\rightarrow {3/2}^+_1$ & 27 & 19.9(8) & - & - \\
${7/2}^+_1\rightarrow {5/2}^+_1$ & 16 & 12.8(12) & 0.0020 & 0.0032(3) \\
\hline
\multirow{2}{*}{} & \multicolumn{2}{c}{$Q_J$ ($e$b)} &
 \multicolumn{2}{c}{$\mu_J$ ($\mu_N^2$)} \\
\cline{2-3} 
\cline{4-5}
          & Th.         & Expt.    & Th.         & Expt.     \\
\hline
${3/2}^+_1$ & +0.475  & +0.160(3) & +0.769 & +0.837943(17) \\
${11/2}^-_1$ & +1.13 & +0.98(8) & -1.161  & -1.001(15) \\
\hline\hline
\end{tabular}
\end{center}
\end{table}


\begin{figure}[htb!]
\begin{center}
\includegraphics[width=\linewidth]{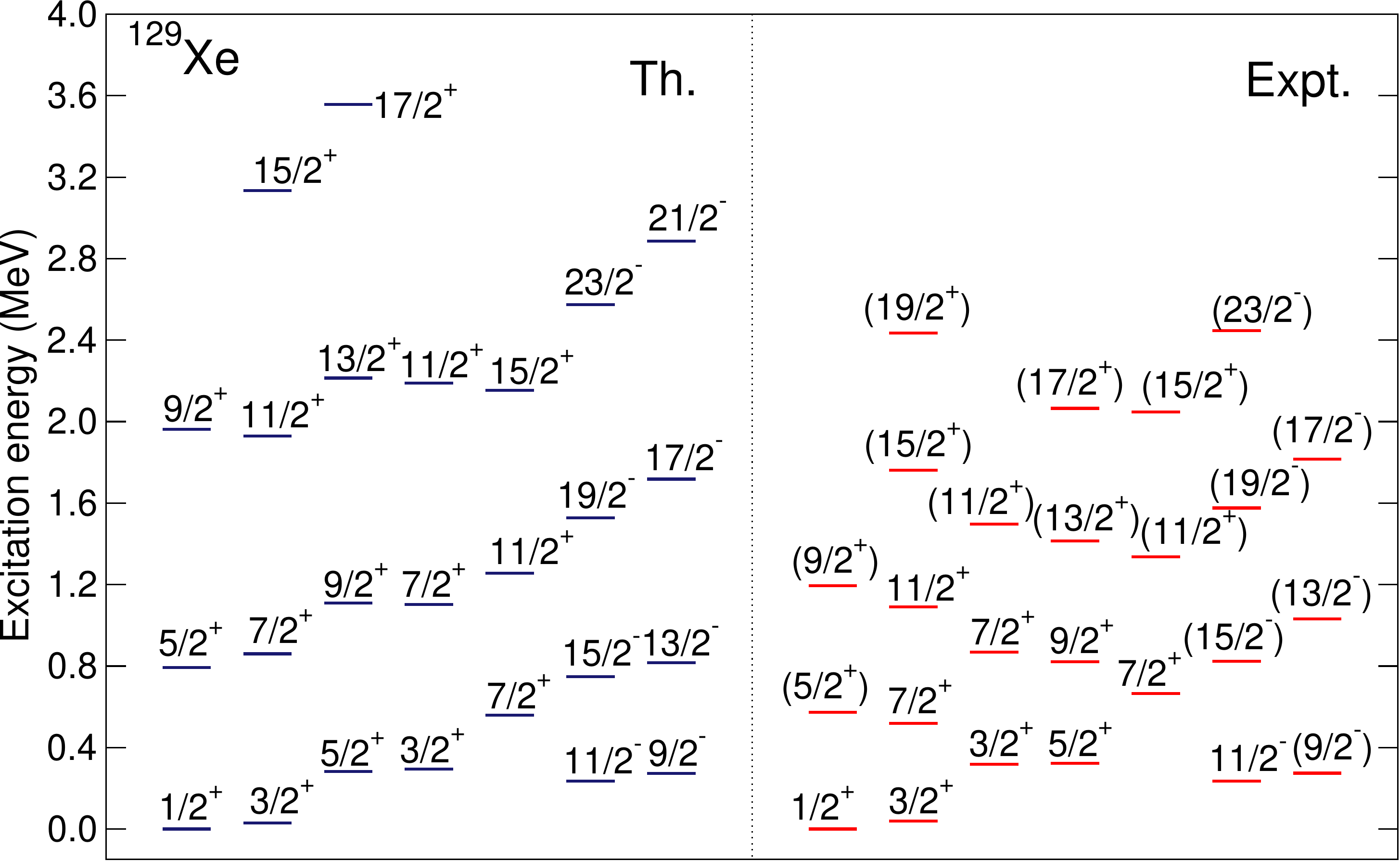}
\caption{(Color online) Comparison between the IBFM theoretical and
 experimental \cite{data} lowest-lying positive- and negative-parity bands of 
 $^{129}$Xe.}
\label{fig:xe129}
\end{center}
\end{figure}

In Fig.~\ref{fig:xe129} we display a detailed comparison between the IBFM theoretical and
 experimental \cite{data} lowest-lying positive- and negative-parity bands of 
$^{129}$Xe. For both parities the present calculation reproduces the structure of the experimental
bands, especially the band-head energies. The low-energy positive-
and negative-parity bands, both theoretical and experimental, exhibit a $\Delta J=2$ systematics
characteristic of the weak coupling limit. The theoretical positive-parity bands are generally more
stretched than the experimental ones, whereas a very good agreement between theory and 
experiment is obtained for the two negative-parity bands.  
Table~\ref{tab:xe129} compares the calculated and experimental
$B(E2)$ and $B(M1)$ values, as well as the electromagnetic
moments of $^{129}$Xe.

\begin{table}[htb]
\caption{\label{tab:xe129}%
Same as in the caption to Tab.~\ref{tab:ba135}, but for $^{129}$Xe. }
\begin{center}
\begin{tabular}{p{2.0cm}cccc}
\hline\hline
\multirow{2}{*}{} & \multicolumn{2}{c}{$B(E2)$ (W.u.)} &
 \multicolumn{2}{c}{$B(M1)$ (W.u.)} \\
\cline{2-3} 
\cline{4-5}
          & Th.         & Expt.    & Th.         & Expt.     \\
\hline
${1/2}^+_{2}\rightarrow {1/2}^+_{1}$ & - & - & 0.010 & 0.0016(5) \\
${1/2}^+_{2}\rightarrow {3/2}^+_{1}$ & 22 & 6.7(23) & 0.049 &
 0.0039(13) \\
${1/2}^+_{2}\rightarrow {3/2}^+_{2}$ & - & - & 0.0039 & 0.0015(5) \\
${1/2}^+_{2}\rightarrow {5/2}^+_{1}$ & 0.018 & 1.4(6) & - & - \\
${3/2}^+_{1}\rightarrow {1/2}^+_{1}$ & 0.89 & 9(4) & 0.0019 & 0.0281(7) \\
${3/2}^+_{2}\rightarrow {1/2}^+_{1}$ & 33 & 23$^{+25}_{-23}$ & - &
 - \\
${3/2}^+_{2}\rightarrow {3/2}^+_{1}$ & 16 & 17$^{+27}_{-17}$ & 0.00091 & 0.003$^{+4}_{-3}$ \\
${3/2}^+_{3}\rightarrow {1/2}^+_{1}$ & 7.4 & $>$0.2 & 0.016 & $>$0.0001 \\
${3/2}^+_{3}\rightarrow {1/2}^+_{2}$ & 8.1 & $>$5.9 & 0.017 & $>$0.0026 \\
${3/2}^+_{3}\rightarrow {3/2}^+_{1}$ & 9.2 & $>$1.6 & 0.0027 & $>$0.00071 \\
${3/2}^+_{3}\rightarrow {3/2}^+_{2}$ & 0.16 & $>$3.4 & 0.029 & $>$0.00037 \\
${3/2}^+_{3}\rightarrow {5/2}^+_{1}$ & 0.25 & $>$4.6 & 0.00054 & $>$0.0005 \\
${5/2}^+_{1}\rightarrow {1/2}^+_{1}$ & 13 & 21(4) & - & - \\
${5/2}^+_{1}\rightarrow {3/2}^+_{1}$ & 46 & 5${\times}10^1$(4) &
 0.0013 & 0.011(5) \\
${5/2}^+_{4}\rightarrow {1/2}^+_{1}$ & 2.0 & 15.4(19) & - & - \\
\hline
\multirow{2}{*}{} & \multicolumn{2}{c}{$Q_J$ ($e$b)} &
 \multicolumn{2}{c}{$\mu_J$ ($\mu_N^2$)} \\
\cline{2-3} 
\cline{4-5}
          & Th.         & Expt.    & Th.         & Expt.     \\
\hline
${1/2}^+_1$ & -  & - & -1.126 & -0.7779763(84) \\
${3/2}^+_1$ & +0.362  & -0.393(10) & +0.72 & +0.58(8) \\
${11/2}^-_1$ & +0.092 & +0.63(2) & -1.247  & -0.891223(4) \\
\hline\hline
\end{tabular}
\end{center}
\end{table}


\begin{figure}[htb!]
\begin{center}
\includegraphics[width=\linewidth]{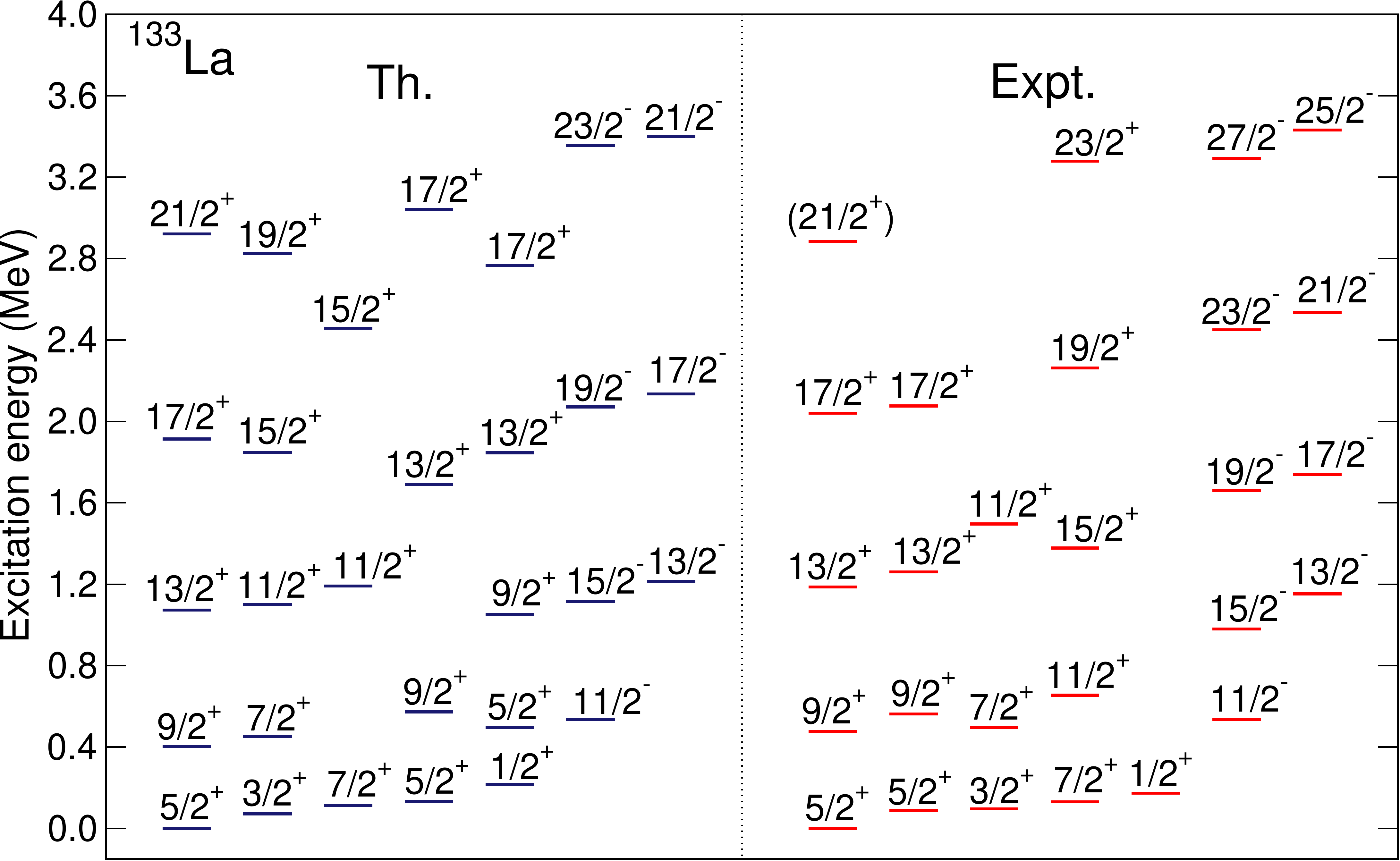}
\caption{(Color online) Same as in the caption to Fig.~\ref{fig:xe129}, but for 
 $^{133}$La.}
\label{fig:la133}
\end{center}
\end{figure}

Next we consider the two odd-Z nuclei, for which the low-lying states 
predominantly correspond to the $1g_{7/2}$, $2d_{5/2}$ (positive-parity) and
$1h_{11/2}$ (negative-parity) proton configurations. 
Figure~\ref{fig:la133} compares several calculated low-energy
positive- and negative-parity bands of $^{133}$La with available
data. One notices a good agreement between the theoretical
and experimental excitation spectra, except for that fact that some of
the calculated bands, that is, the band built on the ${7/2}^+_1$ state and the
two negative-parity bands, appear more stretched than their experimental
counterparts. Similar to $^{129}$Xe, all the low-energy positive- and
negative-parity bands shown here exhibit a $\Delta J=2$ weak-coupling
structure. The calculated and experimental $B(E2)$ and $B(M1)$
values, as well as the electromagnetic moments are listed 
in Tab.~\ref{tab:la133}. 

\begin{table}[htb]
\caption{\label{tab:la133}%
Same as in the caption to Tab.~\ref{tab:ba135}, but for $^{133}$La. }
\begin{center}
\begin{tabular}{p{2.0cm}cccc}
\hline\hline
\multirow{2}{*}{} & \multicolumn{2}{c}{$B(E2)$ (W.u.)} &
 \multicolumn{2}{c}{$B(M1)$ (W.u.)} \\
\cline{2-3} 
\cline{4-5}
          & Th.         & Expt.    & Th.         & Expt.     \\
\hline
${1/2}^+_{1}\rightarrow {3/2}^+_{1}$ & 9.4 & 6(3) & 0.77 & 0.017(6) \\
${1/2}^+_{1}\rightarrow {5/2}^+_{1}$ & 30 & 0.8(3) & - & - \\
${3/2}^+_{1}\rightarrow {5/2}^+_{1}$ & 26 & $>$35 & 0.13 & $>$0.026 \\
${5/2}^+_{2}\rightarrow {5/2}^+_{1}$ & 15 & 2.1(10) & 0.13 & 0.0097(8) \\
${7/2}^+_{1}\rightarrow {5/2}^+_{1}$ & 18 & 11(4) & 0.00011 & 0.0052(9) \\
${7/2}^+_{1}\rightarrow {5/2}^+_{2}$ & 21 & 6.1(20) & 1.0$\times 10^{-5}$ & 0.00068(16) \\
\hline
\multirow{2}{*}{} & \multicolumn{2}{c}{$Q_J$ ($e$b)} &
 \multicolumn{2}{c}{$\mu_J$ ($\mu_N^2$)} \\
\cline{2-3} 
\cline{4-5}
          & Th.         & Expt.    & Th.         & Expt.     \\
\hline
${11/2}^-_1$ & - & - & +6.9 & +7.5(4) \\
\hline\hline
\end{tabular}
\end{center}
\end{table}


\begin{figure}[htb!]
\begin{center}
\includegraphics[width=\linewidth]{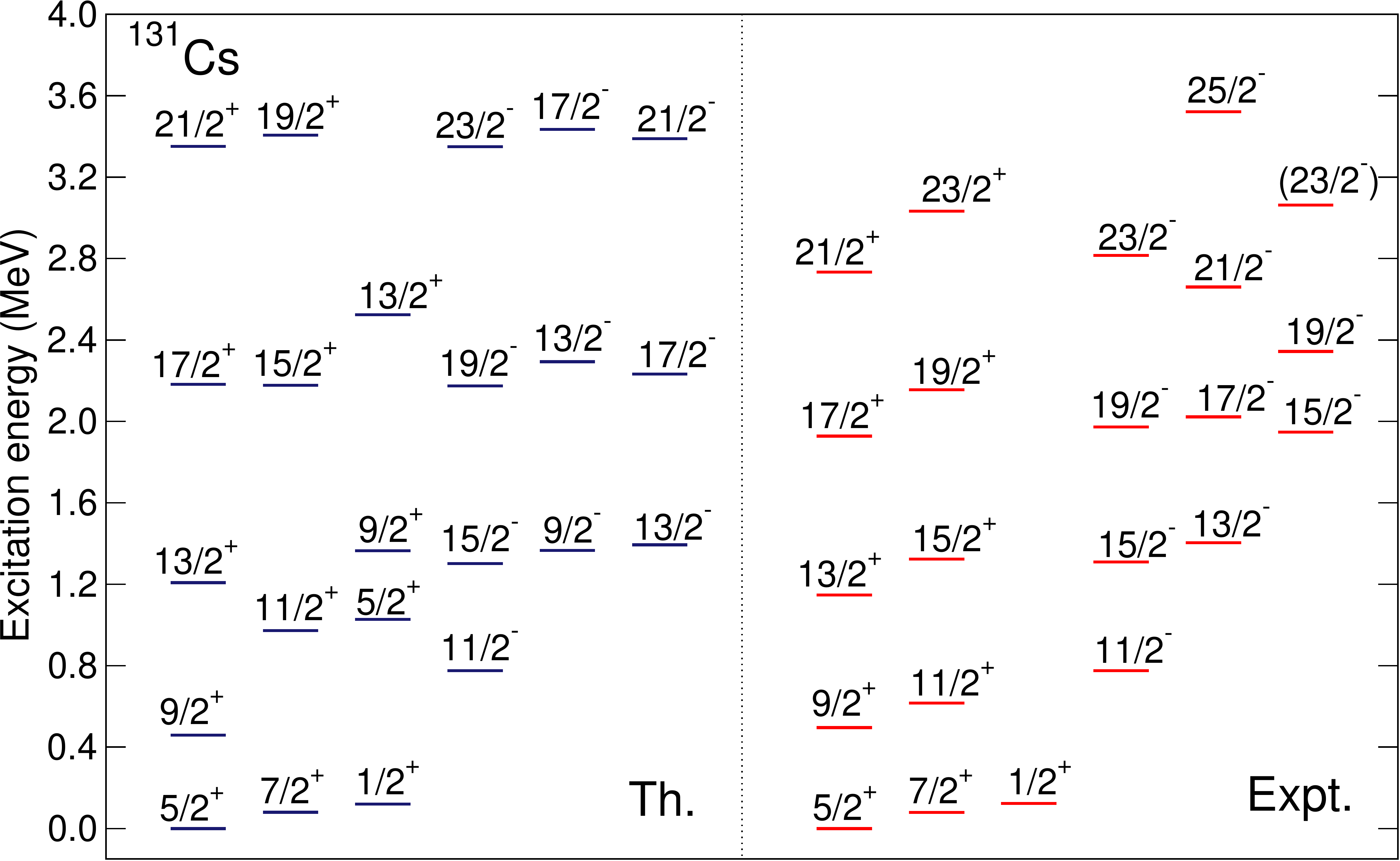}
\caption{(Color online) Same as in the caption to Fig.~\ref{fig:xe129}, but for 
 $^{131}$Cs.}
\label{fig:cs131}
\end{center}
\end{figure}

The theoretical excitation spectrum of $^{131}$Cs, shown in
Fig,~\ref{fig:cs131}, is very similar to that of $^{133}$La and, 
again, a very good agreement is obtained between the IBFM results and experiment. 
The calculated E2 and M1 transition strengths and electromagnetic
moments are compared with the data \cite{data} in 
Tab.~\ref{tab:cs131}. We note that the model calculation qualitatively reproduces 
the complex transition pattern, but obviously the theoretical wave functions do not 
reflect the full extent of configuration mixing in this nucleus.

\begin{table}[htb]
\caption{\label{tab:cs131}%
Same as in the caption to Tab.~\ref{tab:ba135}, but for $^{131}$Cs. Note that the sign is
not known for the experimental $Q_{{5/2}^+_2}$ and $\mu_{{11/2}^-_1}$.}
\begin{center}
\begin{tabular}{p{2.0cm}cccc}
\hline\hline
\multirow{2}{*}{} & \multicolumn{2}{c}{$B(E2)$ (W.u.)} &
 \multicolumn{2}{c}{$B(M1)$ (W.u.)} \\
\cline{2-3} 
\cline{4-5}
          & Th.         & Expt.    & Th.         & Expt.     \\
\hline
${1/2}^+_{1}\rightarrow {5/2}^+_{1}$ & 59 & 69.5(14) & - & - \\
${1/2}^+_{2}\rightarrow {1/2}^+_{1}$ & - & - & 0.013 & 0.0010613(4) \\
${1/2}^+_{2}\rightarrow {3/2}^+_{1}$ & 1.7 & 0.09(4) & 0.011 & $3.4\times 10^{-5}(10)$ \\
${1/2}^+_{2}\rightarrow {3/2}^+_{2}$ & 38 & $>$0.62 & 0.0064 & $>5.8\times 10^{-5}$ \\
${1/2}^+_{2}\rightarrow {5/2}^+_{1}$ & 0.19 & 0.028248(4) & - & - \\
${1/2}^+_{2}\rightarrow {5/2}^+_{2}$ & 4.7 & 0.13835(5) & - & - \\
${3/2}^+_{1}\rightarrow {1/2}^+_{1}$ & 18 & 9(5) & 0.30 & 0.00339(10) \\
${3/2}^+_{1}\rightarrow {5/2}^+_{1}$ & 12 & 0.6(6) & 0.22 & 0.00922(5) \\
${3/2}^+_{1}\rightarrow {5/2}^+_{2}$ & 0.27 & $>$3.9 & 0.0022 & $>4.1\time 10^{-5}$ \\
${3/2}^+_{1}\rightarrow {7/2}^+_{1}$ & 1.4 & 2.36(3) & - & - \\
${3/2}^+_{2}\rightarrow {1/2}^+_{1}$ & 0.94 & 2.4(4) & 0.018 & 0.00057(4) \\
${3/2}^+_{2}\rightarrow {3/2}^+_{1}$ & 0.012 & $>$2.1 & 2.7$\times 10^{-5}$ & $>7.8\times 10^{-5}$ \\
${3/2}^+_{2}\rightarrow {5/2}^+_{1}$ & 0.55 & 2.4(9) & 0.0012 & 0.00064(20) \\
${3/2}^+_{2}\rightarrow {5/2}^+_{2}$ & 0.63 & 0.5(4) & 0.0014 & 0.00071(4) \\
${3/2}^+_{2}\rightarrow {7/2}^+_{1}$ & 25 & 0.2122(3) & - & - \\
${5/2}^+_{2}\rightarrow {5/2}^+_{1}$ & 0.016 & 3.5(3) & 0.0036 & 0.000369(17) \\
${5/2}^+_{2}\rightarrow {7/2}^+_{1}$ & 45 & $>$62 & 3.1$\times 10^{-5}$ & $<2.1\times 10^{-5}$ \\
${7/2}^+_{1}\rightarrow {5/2}^+_{1}$ & 0.10 & 0.64(24) & 0.0010 & 0.00170(5) \\
\hline
\multirow{2}{*}{} & \multicolumn{2}{c}{$Q_J$ ($e$b)} &
 \multicolumn{2}{c}{$\mu_J$ ($\mu_N^2$)} \\
\cline{2-3} 
\cline{4-5}
          & Th.         & Expt.    & Th.         & Expt.     \\
\hline
${5/2}^+_{1}$ & -0.772 & -0.575(6) & +3.42 & +3.543(2) \\
${5/2}^+_{2}$ & +0.370 & 0.022(2) & +0.37 & +1.86(8) \\
${11/2}^-_{1}$ & - & - & +6.9 & 6.3(9) \\
\hline\hline
\end{tabular}
\end{center}
\end{table}

\subsection{Effective $\beta$ and $\gamma$ deformations}

\begin{figure}[htb!]
\begin{center}
\includegraphics[width=0.7\linewidth]{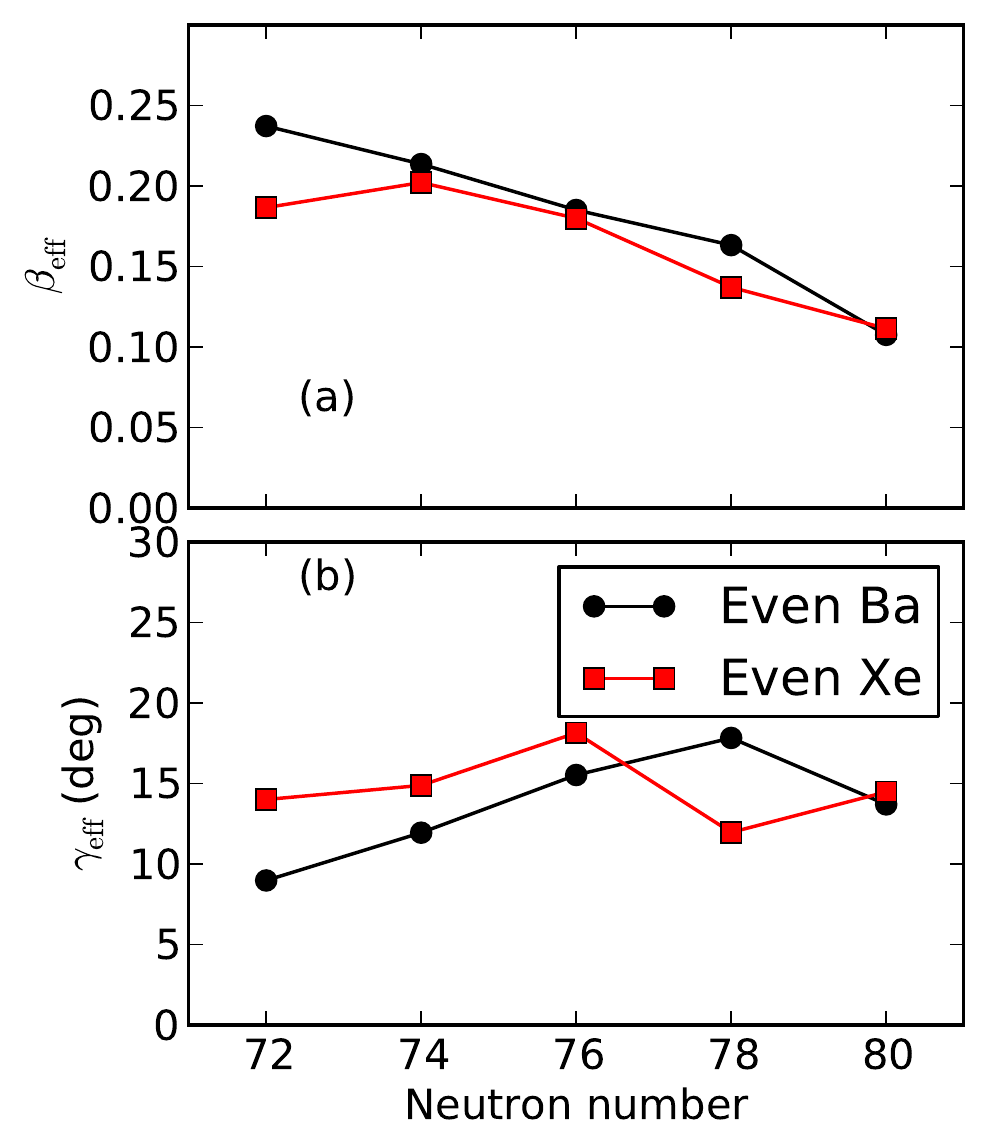}
\caption{(Color online) $\beta_{\rm eff}$ and $\gamma_{\rm eff}$ for
 the $0^+_1$ state of the even-even $^{128-136}$Ba and $^{126-134}$Xe
 isotopes, obtained from the computed q-invariants.} 
\label{fig:even-qinvar}
\end{center}
\end{figure}

\begin{figure}[htb!]
\begin{center}
\includegraphics[width=\linewidth]{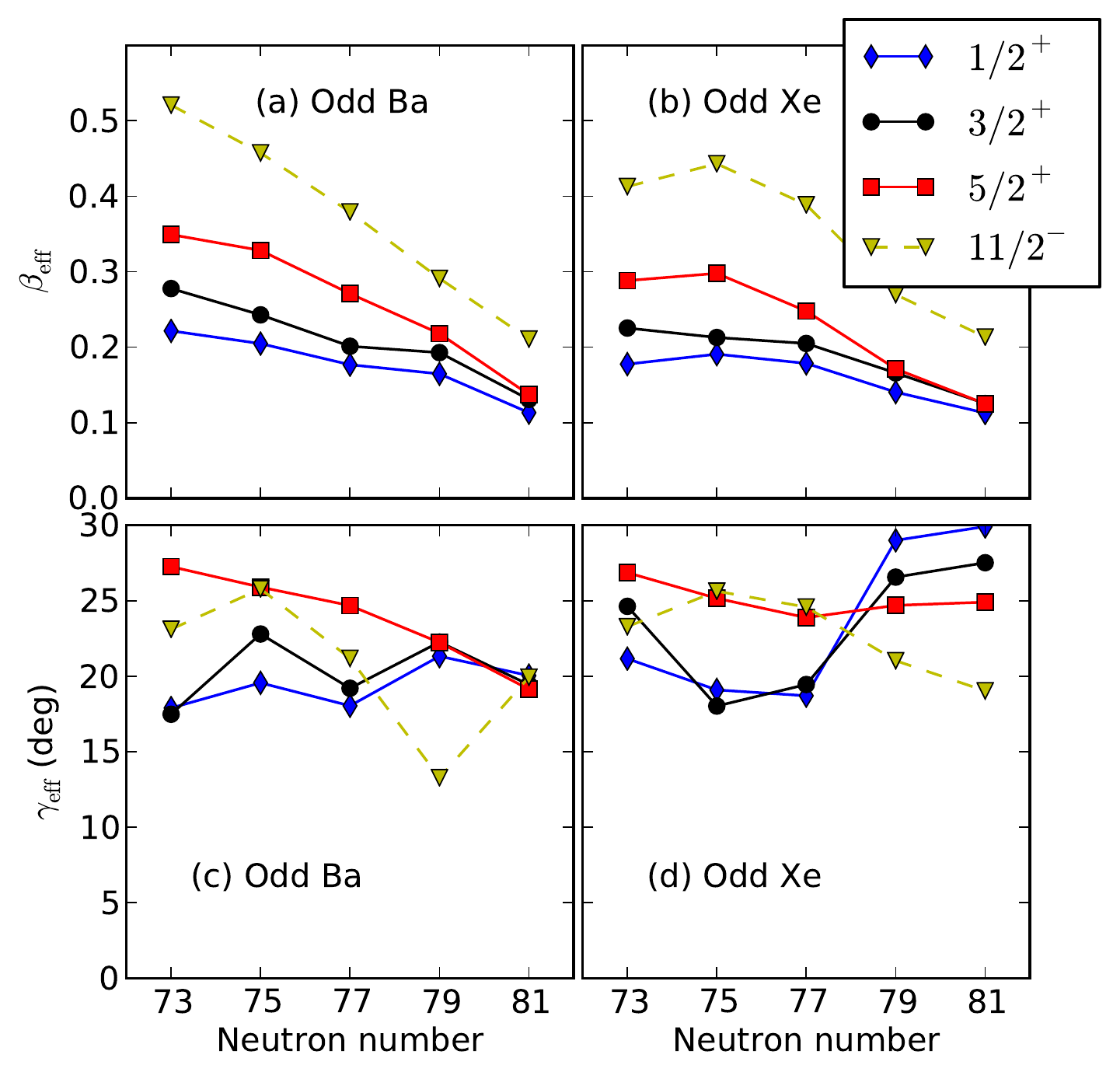}
\caption{(Color online) Same as in the caption to Fig.~\ref{fig:even-qinvar}, but for the
 ${1/2}^+_1$, ${3/2}^+_1$, ${5/2}^+_1$ and ${11/2}^-_1$ states of the
 odd-N $^{129-137}$Ba and $^{127-135}$Xe isotopes.}
\label{fig:oddn-qinvar}
\end{center}
\end{figure}

\begin{figure}[htb!]
\begin{center}
\includegraphics[width=\linewidth]{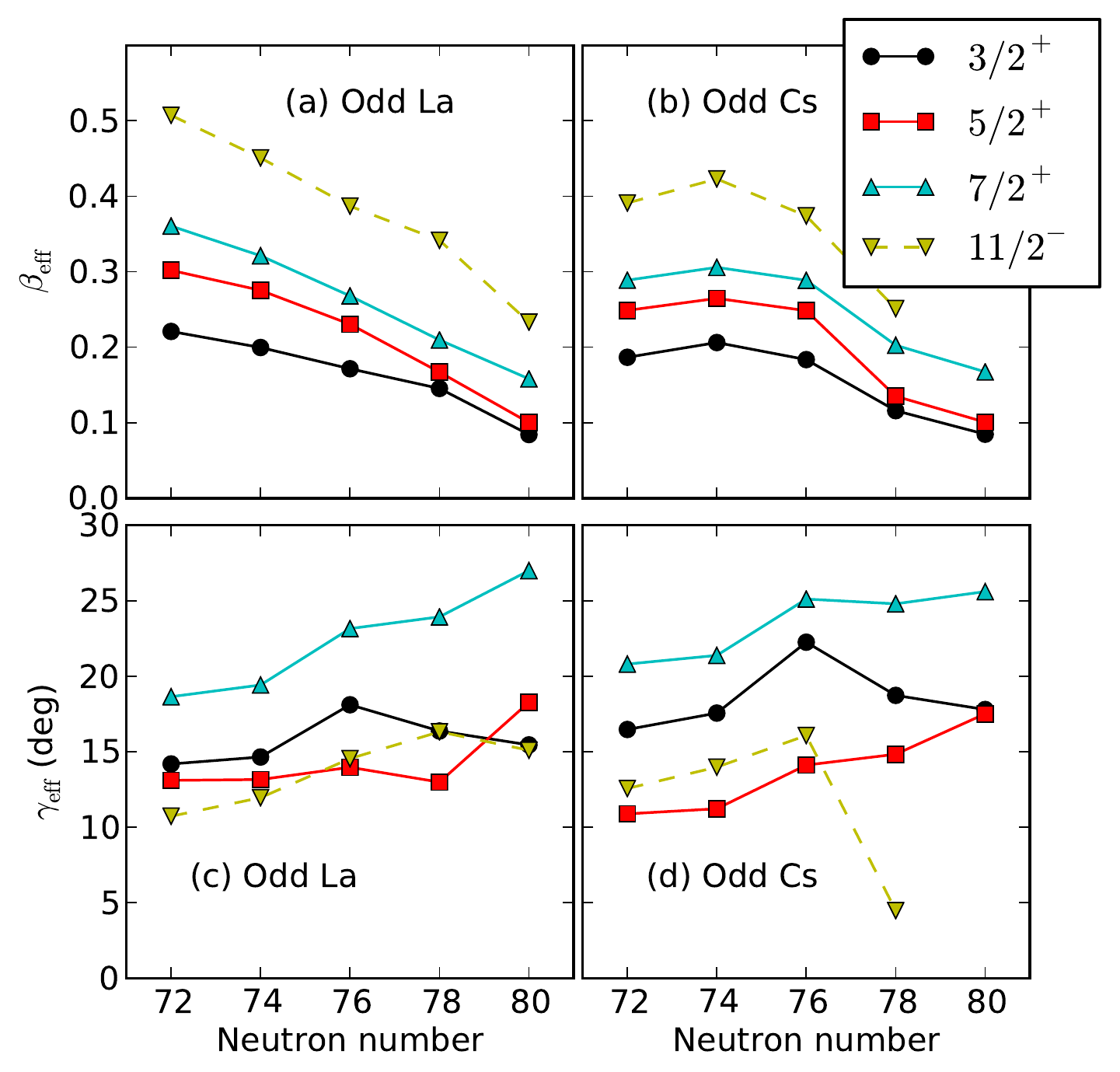}
\caption{(Color online) Same as in the caption to Fig.~\ref{fig:even-qinvar}, but for the
 ${3/2}^+_1$, ${5/2}^+_1$, ${7/2}^+_1$ and ${11/2}^-_1$ states of the
 odd-Z $^{129-137}$La and $^{127-135}$Cs isotopes.}
\label{fig:oddp-qinvar}
\end{center}
\end{figure}

Another signature of possible shape-phase transitions related to the 
$\gamma$-softness of the effective nuclear potential, 
can be computed from $E2$ transition rates. 
Here we specifically analyze quadrupole shape invariants 
\cite{cline1986} (denoted hereafter as q-invariants), 
calculated using $E2$ matrix elements.  
The lowest-order q-invariants for a given state with spin $J$, relevant for the
present study, are defined by the following relations \cite{werner2000}: 
\begin{eqnarray}
q_2=\sum_{i}^n\langle J||\hat Q||J^{\prime}_i\rangle\langle
  J^{\prime}_i||\hat Q||J\rangle
\end{eqnarray}
\begin{eqnarray}
q_3=\sqrt{\frac{7}{10}}|
\sum_{i,j}^{n}
\langle J||\hat Q||J^{\prime}_i\rangle
\langle J^{\prime}_i||\hat Q||J^{\prime}_j\rangle
\langle J^{\prime}_j||\hat Q||J\rangle |
\end{eqnarray}
where $J^{\prime}=J+2$, and the sum is in order of increasing excitation
energies of the levels $J^{\prime}$. 
Only a few lowest transitions contribute to the q-invariants significantly and, in the
present study, the sum runs up to $n=5$. 
For even-even systems, we calculate the q-invariants for the
$0^+_1$ ground state, which means $J=0^+_1$ and $J^{\prime}=2^+$. 
The effective deformation parameters, denoted as $\beta_{\rm eff}$ and $\gamma_{\rm 
eff}$, can be obtained from $q_2$ and $q_3$ \cite{werner2000}: 
\begin{eqnarray}
\label{eq:eff_beta}
 &&\beta_{\rm eff} = \frac{4\pi}{3ZR^2_0}\sqrt{\frac{q_2}{2J^{\prime}+1}(J^{\prime}2J0|JJ)^{-2}} \\
\label{eq:eff_gamma}
 &&\gamma_{\rm eff} = \frac{1}{3}\arccos{\frac{q_3}{q_2^{3/2}}}
\end{eqnarray}
where $R_0=1.2A^{1/3}$ fm, and $(J^{\prime}2J0|JJ)$ is the Clebsch-Gordan
coefficient.

In Fig.~\ref{fig:even-qinvar} we plot $\beta_{\rm eff}$
and $\gamma_{\rm eff}$ for the even-even isotopes $^{128-136}$Ba and 
$^{126-134}$Xe, as functions of the neutron number. 
One notices that, both for Ba and Xe nuclei, $\beta_{\rm
eff}$ exhibits only a gradual decrease with neutron number. 
This correlates with the mean-field result, which indicates that the $\beta$ deformation
does not change significantly as a function of neutron number 
(cf. Figures \ref{fig:evenba-pes} and \ref{fig:evenxe-pes}). 
In contrast, $\gamma_{\rm eff}$
displays a distinct peak at $N=78$ for Ba and at $N=76$ for Xe, which
could be associated with the phase transition between nearly spherical
and prominently $\gamma$-soft shapes. Indeed, the deformation
energy surface at around these neutron numbers resembles the potential
in the E(5) model, which is flat-bottomed in an interval of the axial deformation $\beta$, 
and independent of $\gamma$. 

In the case of odd-A nuclei the spin of the ground state is not
always the same for all isotopes, and we have thus calculated
$\beta_{\rm eff}$ and $\gamma_{\rm eff}$ for several low-lying state. 
Figure~\ref{fig:oddn-qinvar} displays $\beta_{\rm eff}$ and $\gamma_{\rm eff}$
of the states ${1/2}^+_1$, ${3/2}^+_1$, ${5/2}^+_1$ and
${11/2}^-_1$ for the odd-N systems, that is, $^{127-135}$Xe and $^{129-137}$Ba.
Similarly to the corresponding even-even core nuclei, in the 
odd-A Ba nuclei shown in Fig.~\ref{fig:oddn-qinvar} (a) for all four states $\beta_{\rm eff}$ exhibits only a gradual
decrease with the neutron number. For the states ${5/2}^+_1$ and ${11/2}^-_1$ in the
odd-Xe nuclei, however, $\beta_{\rm eff}$ indicates a discontinuity at $N=75$. 
$\gamma_{\rm eff}$, shown in 
Figs.~\ref{fig:oddn-qinvar} (c) and (d), exhibits a significant change 
(either increase or decrease) for many states at $N=79$. 
In addition, $\gamma_{\rm eff}$ for the states ${3/2}^+_1$ and ${5/2}^+_1$
of odd-A Ba nuclei displays another  variation at $N=75$. 
Similar results are also obtained for the odd-Z La and Cs nuclei, as shown in Fig.~\ref{fig:oddp-qinvar}.  
However, $\gamma_{\rm eff}$ for the odd-Z La and Cs
isotopes exhibits a more pronounced signature of shape phase transition when 
compared to the odd-N Ba and Xe nuclei: a significant change at  $N=76$
or 78 for the odd-A La, and at $N=76$ for the odd-A Cs isotopes.


\section{Concluding remarks\label{sec:summary}}


Using the recently proposed method of Ref.~\cite{nomura2016odd}, based on
the microscopic framework of nuclear energy density functionals and the 
particle-core coupling scheme, we have analyzed signatures of QPTs in  
$\gamma$-soft odd-mass nuclei with mass $A\approx 130$.  
The deformation energy surface of the even-even core nuclei, and the spherical
single-particle energies and occupation probabilities of the unpaired nucleon, 
are obtained by relativistic Hartree-Bogoliubov SCMF calculations 
with a specific choice of the energy density functional and pairing
interaction.

The microscopic SCMF calculations determine the parameters of 
boson and  fermion Hamiltonians used to model spectroscopic properties of
the odd-A $^{129-137}$Ba, $^{127-135}$Xe, $^{129-137}$La and
$^{127-135}$Cs nuclei, whereas the strength parameters of the
particle-core coupling are adjusted to reproduce selected empirical 
results of low-energy spectra in odd-A systems. 
The method provides a very good description of spectroscopic
properties of the $\gamma$-soft odd-mass systems. 
Even though phase transitions are smoothed out 
in finite systems, especially a second-order QPT as the one 
considered here, and the physical control parameter takes only integer values (the nucleon
number), the SCMF deformation energy surfaces and the resulting excitation
spectra consistently point to a shape phase transition in the interval $N=76 - 78$, both
in even-even and odd-mass systems. In particular, $\gamma_{\rm eff}$, evaluated using  
E2 matrix elements for transitions between low-lying states, 
clearly exhibits a discontinuity near $N=76$ and 78,
which signals the occurrence of a phase transition between nearly spherical and
$\gamma$-soft shapes. 
The results obtained in this work, as well as in our previous studies on odd-A Sm and Eu
\cite{nomura2016odd,nomura2016qpt}, have shown that the method of
\cite{nomura2016odd} works not only in axially-deformed nuclei, but also in 
$\gamma$-soft or axially-asymmetric odd-mass systems, and enables 
a systematic investigation of the structural evolution in odd-A 
nuclei in medium-heavy and heavy-mass regions. 

The necessity to fit the strength parameters of the boson-fermion coupling Hamiltonian to 
spectroscopic data in the considered odd-mass nuclei, presents a serious limitation of the 
current implementation of our IBF method. In contrast to the parameters of the boson and 
fermion Hamiltonians that are completely determined by the choice of a global EDF and 
pairing interaction, the boson-fermion coupling must be specifically adjusted for each 
odd-mass nucleus. This procedure, of course, limits the applicability to those nuclei 
for which enough low-energy structure data are available to completely determine the 
strength of the various boson-fermion interaction terms. Therefore an important step forward  
would be to develop a method to microscopically determine, or at least constrain, the values 
of the boson-fermion coupling parameters. One possibility would be to perform SCMF calculations for 
odd-A systems and map the resulting deformation energy surface onto the expectation value of the 
IBFM Hamiltonian in the boson-fermion condensate state
\cite{leviatan1988}. SCMF calculations for odd-A nuclei are, 
of course, computationally very challenging and such an approach would be difficult to apply in  
systematic studies of a large number of nuclei. Another strategy would be to derive the boson-fermion 
coupling from a microscopic shell-model interaction between nucleons in
a given valence space \cite{IBFM-Book-scholten}. In 
this approach the parameters can be determined by equating the matrix elements in the IBFM space to 
those in the shell-model space. The disadvantage of this method is that it requires the explicit 
introduction of a new building block, that is, the shell-model interaction. This is certainly an interesting 
problem and will be the topic of future studies and development of the semi-phenomenological model employed 
in the present analysis.

\acknowledgments
K.N. acknowledges support from the Japan 
Society for the Promotion of Science. This work has been supported in part by 
the Croatian Science Foundation -- project ``Structure and Dynamics
of Exotic Femtosystems'' (IP-2014-09-9159) and the QuantiXLie Centre of
Excellence.

\bibliography{refs}

\end{document}